\documentclass[a4paper,11pt]{article}
\usepackage[top=1in, bottom=1in, left=1in, right=1in]{geometry}
\usepackage[UKenglish]{babel}
\usepackage[UKenglish]{isodate}\usepackage[utf8x]{inputenc}
\usepackage{amsmath}
\usepackage{float}
\usepackage{amssymb} 
\usepackage{rotating}
\usepackage{dsfont}
\DeclareMathOperator*{\argmax}{arg\,max}
\usepackage[retainorgcmds]{IEEEtrantools}
\usepackage{graphicx}
\usepackage{tabularx}
\usepackage{subfig}
\usepackage{microtype} 
\usepackage{booktabs}   
\usepackage{bm}         
\usepackage{listings}   
\usepackage{verbatim}   
\usepackage{colortbl}  
\usepackage[dvipsnames]{xcolor}
\usepackage{hyperref}
\hypersetup{
	colorlinks = true,
	citecolor = PineGreen,
	linkbordercolor = {white},
}
\usepackage[colorinlistoftodos]{todonotes}
\usepackage{natbib}
\usepackage{cite}
\usepackage{adjustbox}
\usepackage[titletoc, title]{appendix}
\usepackage{titlesec}
\usepackage{mathtools}
\usepackage{setspace}
\usepackage{rotating}
\titleformat*{\subsubsection}{\itshape}
\usepackage{arydshln}

\begin{document}
	
	\date{}
	\title{
		\vspace{-0pt}
		\textbf{Asymmetric uncertainty : Nowcasting using skewness in real-time data}\\
		\vspace{-10pt} 
	}
\author{Paul Labonne\footnote{Centre for Applied Macroeconomics and Commodity Prices, BI Norwegian Business School. Address: BI Norwegian Business School, Nydalsveien 37, N-0484 Oslo. Email: labonnepaul@gmail.com. This paper has been accepted for publication in the International Journal of Forecasting with DOI \href{https://10.1016/j.ijforecast.2024.05.003}{https://10.1016/j.ijforecast.2024.05.003}. The R and C++ code for replication, as well as the data, are available at: \href{https://github.com/paullabonne/IJF_24}{https://github.com/paullabonne/IJF\_24}. A Python/JAX implementation is available at: \href{https://github.com/paullabonne/scoreJAX}{https://github.com/paullabonne/scoreJAX}. This work is part of the research activities at the Centre for Applied Macroeconomics and Commodity Prices (CAMP) at the BI Norwegian Business School.  I am grateful for funding from the Research Council of Norway through the project "NRC 315008: The Corona-crisis, structural change, and macroeconomic policy". I am also grateful to the Economic Statistics Centre of Excellence (ESCoE) where this work was initiated. I thank Martin Weale, Ivan Petrella, Jamie Cross, Leif Anders Thorsrud, Michele Piffer, Francesca Monti, Ron Smith, and David Kohns for their comments. This paper has also benefited from discussions at the 11th ECB Conference on Forecasting Techniques, International Symposium on Forecasting 2021, 3rd QMUL Economics and Finance Workshop for PhD Post-doctoral Students, Orlando SNDE 2022, Monash University and Sydney TSF 2023.}}
	
	\maketitle

		\vspace{-30pt} 
		\begin{abstract}
			\singlespacing
				This paper presents a new way to account for downside and upside risks when producing density nowcasts of GDP growth. The approach relies on modelling location, scale and shape common factors in real-time macroeconomic data. While movements in the location generate shifts in the central part of the predictive density, the scale controls its dispersion (akin to general uncertainty) and the shape its asymmetry, or skewness (akin to downside and upside risks). The empirical application is centred on US GDP growth and the real-time data come from Fred-MD. The results show that there is more to real-time data than their levels or means: their dispersion and asymmetry provide valuable information for nowcasting economic activity. Scale and shape common factors (i) yield more reliable measures of uncertainty and (ii) improve precision when macroeconomic uncertainty is at its peak. \vspace{10pt}
\end{abstract}
	\newpage
	\section{Introduction}
	
	This paper shows how density nowcasts of gross domestic product (GDP) traditionally generated from first moments of real-time data can be improved by exploiting higher moments. Economists have known since \citet{rhodes_1937} that macroeconomic data exhibit strong co-movements which can be formally estimated with a few common factors. \citet{stock_watson_2002a} show how exploiting these common factors yields large gains in forecasting, and this idea has been dominant ever since. \citet{giannone_reichlin_small_2008} formalise this approach for nowcasting - predicting the recent past, present and near future - which has proved a key tool for policymakers and economic agents. But the recent large fluctuations in economic activity have highlighted important limitations in this method\footnote{For instance, the New York Fed suspended its nowcasting model for about two years following the pandemic, see https://www.newyorkfed.org/research/policy/nowcast/overview.html}, most notably when communicating nowcasting uncertainty. This paper illustrates how extending the idea of common factors to conditional dispersion (variance) and shape (asymmetry or skewness) enhances measures of nowcasting uncertainty and improves precision when macroeconomic uncertainty is at its peak.
	
	To exploit higher moments of real-time data, this paper proposes a flexible skewed-t model in which location, dispersion and asymmetry parameters across series follow parallel factor models. The dynamic features are captured with the score driven methods introduced by \citet{harvey._2013} and \citet{creal_koopman_lucas_2013}. Specifically, this paper extends the univariate skewed-t model of \citet{petrella_2020} to real-time common factor analysis. Score driven models are an observation-driven alternative to non-Gaussian state space models. Unlike the latter, however, their likelihood is generally available in closed form which facilitates estimation considerably. They have been applied successfully to economic forecasting problems notably by \citet{delle_monache_petrella_2017}, \citet{creal_schwaab_koopman_lucas_2014}, \citet{gorgi_koopman_li_2019} and \citet{petrella_2020}.  \citet{buccheri2021filtering} provide the tools for extending the use of score driven models to nowcasting. 
	
	Using this novel model featuring scale and shape common factors, this paper makes two main contributions. First, it shows that real-time macroeconomic data exhibit common factors in their conditional dispersion and asymmetry. These two factors carry economic meaning; they represent general macroeconomic uncertainty and risk. Second, it shows that two proven and widely used nowcasting methods (factor-augmented MIDAS and mixed-frequency factor models) can be extended to exploit these new empirical facts. While using scale and shape common factors exogenously in a MIDAS set up is conceptually straightforward, modelling them endogenously in a mixed-frequency model is more challenging. The trick here is to aggregate the real-time data into the frequency of GDP growth to alleviate the frequency mismatch. 
	
	The proposed nowcasting approach featuring scale and shape common factors is used to produce density nowcasts of US GDP growth in real-time using about thirteen real-time series. Estimation is carried out recursively using monthly vintages from Fred-MD \citep{mccracken_fred-md_2016} which facilitates both real-time estimation and replication. The estimation exercise starts in February 2007 and ends with the vintage of April 2023; thus it includes both the financial crisis and the Covid-19 pandemic. To assess the usefulness of scale and shape common factors, the full model is compared with nested specifications where scale and shape dynamics are removed, as well as an autoregressive (AR) model with student t-distributed errors. Each model is evaluated based on two metrics: calibration and sharpness. Calibration evaluates a model's ability to provide reliable estimates of uncertainty; sharpness evaluates how close the probabilistic predictions are to the actual observations.
	
	The real-time estimation exercise elucidates four important empirical findings. First the scale and shape common factors retrieved in real-tiem data carry economic meaning: they are reliable real-time indicators of general macroeconomic uncertainty and risk. Second, these two factors provide a new way to model data which are weakly related to economic activity through their location but are a good indicator of macroeconomic uncertainty, such as stock market data. Third, incorporating them in a nowcasting model can improve calibration and thus the reliability of probabilistic predictions. Fourth, the shape common factor proves to be a key feature to capture accurately the large economic downturn and subsequent recovery during the pandemic.
	
	The rest of the paper is structured as follows. Section \ref{sec:litt_review} provides a literature review that links the proposed framework to established nowcasting methods. Section \ref{sec:model} presents the baseline skew-t score driven model with location, scale and shape common factors. Section \ref{sec:MF} shows how to extend the model to accommodate mixed frequency data. Section \ref{sec:real_time} evaluates the performance of the proposed methodology in real time using a recursive estimation setting. Section \ref{sec:results} discusses the main results while the last section concludes.

	\section{Link to established nowcasting methods}\label{sec:litt_review}
      
  Modelling jointly high-frequency data, such as Fred-MD data, and series available on lower frequencies, such as GDP, can be done in two ways. Either the model is specified at the highest frequency, in which case filtering techniques are employed to model latent high-frequency components in the low-frequency series; or it is specified at the lowest frequency, in which case high-frequency variables are aggregated first before being modelled with low-frequency series. 
  
  When using filtering techniques, the approximation of \citet{Mariano_Murasawa_2003} is employed to relate the latent high-frequency components to low-frequency observations.\footnote{The use of this approximation is necessary because he temporal aggregation constraint facing quarterly flow variables (that is, they are the tree-month sum of their monthly components) becomes non-linear when modelling data in log differences as is commonly done.} High-dimensional data are handled either by modelling a few latent common factors, yielding mixed-frequency dynamic factor models \citep{giannone_reichlin_small_2008}, or by using Bayesian shrinkage methods when using a mixed-frequency vector autoregression framework \citep{schorfheide_song_2015}. These approaches provide a coherent framework for nowcasting and evaluating the benefit of each new data release \citep{handbook_forecasting}. Density nowcasts from these models are improved by modelling time-varying variance parameters \citep{marcellino_porqueddu_venditti_2016, Petrella_2017, Petrella_2020_bis}. 
   
   When the joint modelling occurs at the lowest frequency, real-time data are used as exogenous variables, or regressors. Bridge equations, a popular approach in economic institutions, aggregate regressors before estimation using known functions. An alternative approach due to \citet{ghysels_sinko_valkanov_2007} is mixed-data sampling (MIDAS) which uses estimated lag polynomials for aggregating regressors. These techniques can be extended to a large set of data by either i) using penalised estimation as in \citet{babii_ghysels_striaukas_2021}, ii) using common factors extracted in a first step as regressors \citep{carriero_galvao_kapetanios_2019}, or iii) using regressors one at a time and combining the resulting forecasts as in \citet{kitchen_monaco_2003} and \citet{mazzi_mitchell_montana_2013}.    Regarding density predictions, \citet{pettenuzzo_timmermann_valkanov_2016} model time-varying volatility in a MIDAS set up and \citet{ferrara_mogliani_sahuc_2022} apply MIDAS techniques to quantile regression; see also \citet{carriero_clark_marcellino_2022} for recent developments in this field.
   
   All the models discussed above have one thing in common: they all use real-time data in their levels (or growth rates) or alternatively model their conditional means. But they do not exploit their higher moments. To bridge this shortcoming, the next section proposes a model for capturing co-movements in second and third moments of real-time data, more precisely in their dispersion and asymmetry. Section \ref{sec:MF} then shows how these co-movements, captured with scale and shape common factors, may be exploited in a mixed-frequency setting using either a mixed-frequency factor model or a factor-augmented MIDAS model.
   
    \section{A skew-t score-driven approach for nowcasting}\label{sec:model}
   This section sets out a skew-t model fitted for modelling location, scale, and shape common factors in a nowcasting setting.  Score-driven techniques~\citep{creal_koopman_lucas_2013,harvey._2013} are used to model
   unobserved components, specifically secular trends and common autoregressive components. The model presented below is inspired by the score-driven dynamic factor model of~\citet{creal_schwaab_koopman_lucas_2014}, who model
   a location common factor in data following different conditional distributions. Here, the dynamic factor structure is extended to scale and shape parameters using the score-driven equations of~\citet{petrella_2020}, who
   present a univariate skew-t score-driven model. Finally the model is adapted to a nowcasting setting by adding a filtering equation following the method of~\citet{buccheri2021filtering}.
   
   Real-time monthly series from FRED-MD, $y_{i,t}$, $i = 1, \ldots , N$, $t = 1, \ldots , T$, are modelled as:
   \begin{equation}
   	\label{ref:observation_equation}
   	y_{i,t} = \mu_{i,t} + v_{i,t}, \quad v_{i,t} \sim Skt(0, \sigma_{i,t}, \alpha_{i,t}, \nu_{i}),
   \end{equation}
   where $\mu_{i,t}$ is the location and $v_{i,t}$ the prediction error. The prediction error is assumed to follow a centred skew-t distribution, where $\sigma_{i,t} >0$ is the scale parameter, $\alpha_{i,t} \in (-1, 1)$ is the shape parameter, and $\nu_i>1$ is the tail parameter.
   
   Each time-varying parameter $x_{i,t}$, $x \in \{\mu, \sigma, \alpha\}$, is split between a trend, $\bar{x}_{i,t}$, and a common component, $\tilde{x}_{t}$, such that:
   \begin{equation}
   	\label{ref:link_functions}
   	\begin{split}
   		\mu_{i,t} &= \bar{\mu}_{i,t}+ \Lambda_{\mu,i}\tilde{\mu}_{t},\\
   		\sigma_{i,t} &= \text{exp}(\bar{\sigma}_{i,t}+ \Lambda_{\sigma,i}\tilde{\sigma}_{t}),\\
   		\alpha_{i,t} &= \text{tanh}(\bar{\alpha}_{i,t}+ \Lambda_{\alpha,i}\tilde{\alpha}_{t}),
   	\end{split}
   \end{equation}
   where $\Lambda_{x,i}$ is the factor loading controlling the dependence of series $i$ on the common factor in parameter $x$. The exponential function ensures that scale parameters are positive, and the hyperbolic tangent function is used to constrain the shape parameters to lie in $(-1,1)$. The latent components are governed by a score-driven model taking the form:
   \begin{equation}
   	\label{ref:factor_model}
   	\begin{split}
   		\bar{x}_{i,t+1} &=  \bar{x}_{i,t} + A_{\bar{x},i}  s_t({\bar{x}_{i,t}}),\\
   		\tilde{x}_{t+1} &=  \phi_{x,1} \tilde{x}_{t }+ \phi_{x,2} \tilde{x}_{t-1}+ A_{\tilde{x}}  s_t({\tilde{x}_{t}}).
   	\end{split}
   \end{equation}
   
   ~\citet{Petrella_2017} demonstrate that random-walk specifications for secular trends are robust to discrete breaks. The location common component represents the business cycle, which is
   commonly modelled as an autoregressive process of order two. The scale common component captures common volatility shocks, like the one generated by the Covid-19 pandemic. It is specified as a stationary AR(1) ($\phi_{\sigma,2}$
   = 0). Finally, the shape common component captures simultaneous shifts in the asymmetry of prediction errors, like those typically arising at the beginning of recessions and in the early phases of recovery. It is also
   modelled as a stationary AR(1) ($\phi_{\alpha,2}$ = 0). The initial values of the trend components are estimated, while the common components are initialised with their unconditional means.
   
   Following~\citet{creal_koopman_lucas_2013} and~\citet{harvey._2013}, the dynamic in the model comes from the score $s_t(.)$, the first derivative of the log joint conditional density. The gains, $A_{\bar{x}_i}$ and $A_{\tilde{x}}$, pre-multiply the score and determine the degree of variation in the unobserved components (their dependence on the dynamic introduced by the score). Taking the common component as an example, the
   score at time $t$ takes the form:
   \begin{equation}
   	\label{ref:score_eq}
   	s_t(\tilde{x}_{t}) = \frac{\partial \text{log}p(y_t\vert f_t, \Theta)}{\partial \tilde{x}_{t}},
   \end{equation}
   where the vector $f_t$ contains the dynamic unobserved components $\tilde{x}_{t}$ and $\bar{x}_{i,t}$ and their lags. The vector $\Theta$ contains the static parameters which are estimated, specifically the degrees of freedom $\nu_{i}$, the gains and initial values for the trends $A_{\bar{x}, i}$ and $\bar{x}_{i,0}$, as well as the gains, initial values, persistence parameters and factor loadings linked to the common components $A_{\tilde{x}}$,  $\tilde{x}_{0}$, $\phi_{x,1}$, $\phi_{x,2}$ and $\Lambda_{x,i}$.
   
   The score is often weighted with the information matrix, or its square root, to improve estimation. However, when the dynamic parameters are decomposed into unobserved components, the information matrix often becomes
   singular. While~\citet{creal_schwaab_koopman_lucas_2014} overcome this problem by using the pseudo-inverse, this paper discards the weighting completely, which is a popular strategy when the information matrix is difficult
   to retrieve. The pseudo-inverse is not guaranteed to be continuous, and relying on discontinuous functions creates issues when using gradient-based methods for estimation. Discarding the score weighting also has the
   advantage of being much faster, since it avoids inverting a potentially large matrix at each time iteration. Appendices \ref{ap:sim_dgp} \ref{ap:sim_filter} provide sets of simulations to verify the stability of the resulting
   filter and data generating process.
   
   The data are assumed to be cross-sectionally independent, conditional on past information, so the contemporaneous joint log density is simply the sum of the log marginal densities:
   \begin{equation}
   	\label{eq:log_density_indep}
   	\text{log} p(y_t\vert f_t, \Theta) = \sum^N_{i=1}\delta_{i,t} \text{log} p(y_{i,t}\vert f_t, \Theta),
   \end{equation}
   where $\delta_{i,t}$ is zero if observation $y_{i, t}$ is missing and one otherwise. Following~\citet{petrella_2020}, the form of the skew-t distribution is taken
   from~\citet{gomez_torres_bolfarine_2007}; thus the log density of series $i$ at time $t$ takes the form:
   \begin{equation}
   	\label{eq:loglik-function}
   	 \text{log} p(y_{i,t}\vert f_t, \Theta) = \text{log}C(\eta_i) -\frac{1}{2}\text{log}\sigma^2_{i,t}
   	- \frac{\eta_i+1}{2\eta_i}\text{log}\Bigl[1+\frac{\eta_i(y_{i,t} - \mu_{i,t})^2}{(1-sgn(y_{i,t} - \mu_{i,t})\alpha_{i,t})^2\sigma^2_{i,t}}\Bigr], 
   \end{equation}
   where $C(\eta_i) = \frac{\sqrt{\eta_i}\Gamma\big(\frac{\eta_i+1}{2\eta_i}\big)}{\sqrt{\pi} \Gamma\big(\frac{1}{2\eta_i}\big)}$; $\Gamma(.)$ is the gamma function and $sgn(.)$ is the sign function. The distribution is skewed towards positive values if $\alpha_{i,t}<0$ and towards negative values if $\alpha_{i,t}>0$. When the tail parameters are constrained to be very large and the shape parameter constrained to zero, the distribution is equivalent to a normal distribution. The parameter $\eta_i = 1/\nu_i$ is the inverse tail parameter.
   
   \subsubsection*{Adapting the model to a nowcasting setting}
   
   Unlike state-space models, score-driven models are fully deterministic, conditional on past information. Latent states are predictable perfectly in $t-1$ and there is no room for improvement when new data for time
   $t$ are released; one-step-ahead forecasts are not subject to any uncertainty. In a purely forecasting exercise, that would not be a problem.\footnote{\citet{koopman_lucas_scharth_2016} show that score-driven
   	models and non-Gaussian state-space models have similar predictive accuracy for a wide range of model specifications.} In a nowcasting exercise, however, updating latent states such as the common components, and thus
   nowcasts as well, following the release of contemporaneous information is a critical feature.
   
   \citet{buccheri2021filtering} show that score-driven recursions can be separated between an updating step when latent states are updated using current information, and a predictive step moving the latent states to the next
   period. When using this strategy, the dynamics for the vector of unobserved components $f_t$ can be written as:
   \begin{equation}
   	\label{eq:updating_step}
   	\begin{split}
   		f_{t+1} &= Bf_{*, t},\\
   		f_{*, t} &= f_{t} + A s_t(f_t),
   	\end{split}
   \end{equation}
   where $D$ is a diagonal matrix of gains with only positive elements. Using this specification, nowcasts are produced by using the filtered vector $f_{*, t}$ rather than one-step-ahead predictions $f_{t}$.
   
   Appendix \ref{ap:score} shows the full matrix representation of recursion \eqref{eq:updating_step}, including the scores.
   
   \subsection{Estimation with maximum likelihood}
   \label{sec:estimation}
   The model is estimated using maximum likelihood, such that:
   \begin{equation}
   	\label{eq:MLE}
   	\hat{\Theta} = \argmax_{\Theta} \sum^{T}_{t = 1}\text{log} p(y_t\vert f_t, \Theta).
   \end{equation}
   
   Different permutations of the factor loadings and gain parameters of the autoregressive components ($\Lambda_{x,i}$ and $A_{\tilde{x}}$, $x \in \{\mu, \sigma, \alpha\}$) would yield similar log likelihoods, resulting in identification problems. This
   is a well-known feature of factor models. Here identification is achieved following~\citet{creal_koopman_lucas_2013} by setting the factor loadings of the first series to one. Separately, while the trend is
   series-specific, the unobserved component affects all series. Their scores are therefore not perfectly collinear, which makes identifying each component feasible. 
   
   The log likelihood function is typically maximised using quasi-Newton methods relying on the gradient of the total log-likelihood function with respect to the unknown parameters. This paper makes use of automatic differentiation to retrieve the gradient, which is a highly precise and efficient approach. Specifically, the log likelihood is maximised using the widely used BFGS method available in
   R~\citep{r_core}; the score-driven recursion is written in C++ using Rcpp~\citep{Eddelbuettel2011}; and the gradient is derived with the automatic differentiation library CppAD.
   
   The maximum likelihood approach for estimating score-driven models has been studied thoroughly by \citet{blasques_2020}. But there is no guarantee that location, scale, and shape common factors can be estimated precisely
   with a limited sample.  Appendix \ref{ap:sim_estimation} shows a set of simulations investigating this issue. The average standard deviation of the factor estimates across the simulations is 0.10 for the shape, 0.09 for the location, and 0.21 for the scale.
   
   \section{Mixed-frequency extension}\label{sec:MF}

   This section presents two approaches to exploit the factor model presented in the previous section in a mixed-frequency setting where monthly FRED-MD data are used to nowcast quarterly GDP growth. The first approach
   consists of extending the factor model to accommodate quarterly data by modelling latent monthly components in quarterly figures using the approximation of~\citet{Mariano_Murasawa_2003}. The second approach works in two
   stages. First, the factor model is estimated without GDP growth so that all series remain on a monthly frequency. Second, the resulting estimated monthly location, scale, and shape common factors are aggregated to
   quarterly quantities using MIDAS techniques~\citep{ghysels_sinko_valkanov_2007} and used as regressors for nowcasting GDP growth in a univariate score-driven setting similar to~\citet{petrella_2020}.
   
   Separately, to investigate the potential benefits of modelling scale and shape common factors, it is useful to compare the unrestricted model with versions of the model where scale and shape dynamics are gradually removed. This yields three models: a model with location, scale, and shape common factors, a model with location and scale common factors, and a model with location common factor only. When adding the mixed-frequency extensions, this gives six models. Each model is now clearly set out below.
   
   \subsection{Location models}

   \subsubsection*{Mixed-frequency factor model}
   
   Quarterly GDP growth is added to the FRED-MD data in the vector of observation $y_t$. It shows a quarterly figure in the last month of each quarter that is missing in other months. This is the common way of using
   mixed-frequency data in state-space models~\citep[see][]{handbook_forecasting}, and the strategy is the same with score-driven models. A monthly latent location, $\mu_{1,t}$, is modelled jointly with the location
   parameters of monthly series $\mu_{i,t}$, $i = 2, \ldots , N$, and related to quarterly observations by replacing equation~\eqref{ref:observation_equation} with:
   \begin{align}
   	y_{1,t} &= \frac{1}{3} \mu_{1,t}  + \frac{2}{3}\mu_{1,t-1} + \mu_{1,t-2} + \frac{2}{3}\mu_{1,t-3}+ \frac{1}{3}\mu_{1,t-4} + v_{1,t}, \quad v_{1,t} \sim Skt(0, \sigma_{1,t}, \alpha_{1,t}, \nu_{1}), \label{eq:approximation_g} \\
   	y_{i,t} &= \mu_{i,t} + v_{i,t}, \quad v_{i,t} \sim Skt(0, \sigma_{i,t}, \alpha_{i,t}, \nu_{i}),
   \end{align}
   $i = 2, \ldots , N$, $t = 1, \ldots , T$. While~\citet{Mariano_Murasawa_2003} popularised the use of this approximation,~\citet{mitchell_2005} showed in fact that this is a very precise approximation of the true temporal aggregation constraint facing flow
   variables;  Appendix \ref{ap:loc_disag} details its derivation. Note that when an observation is missing, its associated scores are set to zero.
   
   The link functions \eqref{ref:link_functions} and transition equations~\eqref{ref:factor_model} remain unchanged. The model is estimated with the scale and shape dynamics turned off by fixing $\sigma_{i,t} = \sigma_{i}$ and $\alpha_{i,t} = 0$, $i = 1, \ldots , N$.
   
   \subsubsection*{Factor-augmented MIDAS model}
   
   The factor-augmented MIDAS method works in two stages, where the monthly factor model \eqref{ref:observation_equation}--\eqref{ref:factor_model} is estimated first with $\sigma_{i,t} = \sigma_{i}$ and $\alpha_{i,t} = 0$, $i = 1, \ldots , N$. The filtered estimates of the location common factor $\hat{\tilde{\mu}}_{*,t}$ are retrieved and used in the following univariate model for GDP growth:
   \begin{equation}
   	y_{t} = \mu_{t} + v_{t}, \quad v_{t} \sim Skt(0, \sigma, 0, \nu),\end{equation}
   for $t = q\times3, \quad q = 1,2, \ldots ,T/3$. To keep homogeneous notation across models, the univariate MIDAS model is written at the monthly frequency but the subscript $i$, indicating the series, is dropped. The quarterly observations $y_{t}$ are observable only in the last month of the quarter. The location follows:
   \begin{equation}
   	\label{eq:midas_loc}
   	\begin{split}
   		\mu_{t} &= \bar{\mu}_{t} + \beta_{\mu}\sum^{j=4}_{j=0}B(j,\theta_{\mu})\hat{\tilde{\mu}}_{*, t-j},\\
   		\bar{\mu}_{t+1} &= \bar{\mu}_{t} + A_{\mu}s_{\mu,t},
   	\end{split}
   \end{equation}
   for $ t = q\times3$, $q = 1,2, \ldots ,T/3$. See~\citet{petrella_2020} for details on the scaled score $s_{\mu,t}$ where, unlike in \eqref{ref:score_eq}, a weighting is used. Note that common factors are observable every month, while the trend is updated only in the last month of the quarter.
   $B(h,\theta)$ is the exponential two-parameter Almon lag of~\citet{ghysels_sinko_valkanov_2007} given by:
   \begin{equation}
   	B(j,\theta) = \frac{exp(j.\theta_1 + j^2.\theta_2)}{\sum^{j=4}_{j=0}exp(j.\theta_1 + j^2.\theta_2)}.
   \end{equation}
   The univariate score-driven model is estimated with maximum likelihood similarly to the factor model.
   
   \subsection{Location--scale models}

   \subsubsection*{Mixed-frequency factor model}
   
   While the approximation used in \eqref{eq:approximation_g} works well to model monthly location components in quarterly data, the prediction error remains monthly. This means that the scale parameter of GDP growth
   cannot be modelled directly with the scales of the monthly FRED-MD series. In addition, the approximation of~\citet{Mariano_Murasawa_2003} is only suited to location parameters. However, modelling dependencies across
   series in scale parameters is possible when using a rolling quarterly model. To do so, the real-time data are aggregated into rolling quarterly observations (quarter-on-quarter deviations observed
   monthly).\footnote{\citet{labonne_weale_2020} show that when rolling quarterly series are not subject to important measurement errors, it is possible to interpolate the monthly path very precisely with \eqref{eq:mm}.
   	Hence, there should be little loss of information when using rolling quarterly observations instead of monthly observations.} This introduces serial correlation in the monthly series, which is addressed by using the
   approximation of~\citet{Mariano_Murasawa_2003}. The temporal aggregation strategies for jointly modelling quarterly and monthly observations and modelling rolling quarterly observations are essentially the same; in both
   cases it is necessary to go back to the underlying monthly model. Thus equation \eqref{ref:observation_equation} becomes:
   \begin{align}
   	\label{eq:mm}
   	y_{i,t} =& \frac{1}{3} \mu_{i,t}  + \frac{2}{3}\mu_{i,t-1} + \mu_{i,t-2} + \frac{2}{3}\mu_{i,t-3} + \frac{1}{3}\mu_{i,t-4} + v_{i,t}, \quad \nonumber\\
   	&v_{i,t} \sim Skt(0, \sigma_{i,t}, \alpha_{i,t}, \nu_{i}),
   \end{align}
   this time for all series $i = 1, \ldots , N$, $t = 1, \ldots , T$. The resulting model given by equations~\eqref{eq:mm}, \eqref{ref:link_functions}, and \eqref{ref:factor_model} is estimated with $\alpha_{i,t} = 0$.
   
   \subsubsection*{Factor-augmented {MIDAS} model}
   
   The monthly factor model \eqref{ref:observation_equation}--\eqref{ref:factor_model} is estimated first with $\alpha_{i,t} = 0$, $i = 1, \ldots , N$, to retreive the filtered estimates of the location and scale common factors $\hat{\tilde{\mu}}_{*,t}$ and $\hat{\tilde{\sigma}}_{*,t}$. The univariate model for GDP growth becomes:
   \begin{equation}
   	y_{t} = \mu_{t} + v_{t}, \quad v_{t} \sim Skt(0, \sigma_t, 0, \nu),\end{equation}
   for $t = q\times3, \quad q = 1,2, \ldots ,T/3$. The location follows equation~\eqref{eq:midas_loc}, while a new equation is added for the scale:
   \begin{equation}
   	\label{eq:midas_scale}
   	\begin{split}
   		\sigma_{t} &= \bar{\sigma}_{t} + \beta_{\sigma}\sum^{j=4}_{j=0}B(j,\theta_{\sigma})\hat{\tilde{\sigma}}_{*, t-j},\\
   		\bar{\sigma}_{t+1} &= \bar{\sigma}_{t} + A_{\sigma}s_{\sigma,t},
   	\end{split}
   \end{equation}
   for $ t = q\times3$, $q = 1,2, \ldots ,T/3$. See~\citet{petrella_2020} for details on the scaled score $s_{\sigma,t}$.
   
   \subsection{Location--scale--shape models}
   
   \subsubsection*{Mixed-frequency factor model}
   
   The mixed-frequency factor model featuring location, scale, and shape common factors follows the rolling quarterly specification given by \eqref{eq:mm}, \eqref{ref:link_functions}, and \eqref{ref:factor_model} previously set out but is estimated without any restrictions on the shape parameters.
   
   \subsubsection*{Factor-augmented {MIDAS} model}
   
   The monthly factor model \eqref{ref:observation_equation}--\eqref{ref:factor_model} is first estimated without any restrictions to retreive the filtered estimates of the location, scale and shape common factors $\hat{\tilde{\mu}}_{*,t}$, $\hat{\tilde{\sigma}}_{*,t}$ and $\hat{\tilde{\alpha}}_{*,t}$. The univariate model for GDP growth becomes:
   \begin{equation}
   	y_{t} = \mu_{t} + v_{t}, \quad v_{t} \sim Skt(0, \sigma_t, \alpha_t, \nu),\end{equation}
   for $t = q\times3, \quad q = 1,2, \ldots ,T/3$. The location and scale follow the MIDAS equations~\eqref{eq:midas_loc} and \eqref{eq:midas_scale}, while the shape follows:
   \begin{equation}
   	\begin{split}
   		\alpha_{t} &= \bar{\alpha}_{t} + \beta_{\alpha}\sum^{j=4}_{j=0}B(j,\theta_{\alpha})\hat{\tilde{\alpha}}_{*, t-j},\\
   		\bar{\alpha}_{t+1} &= \bar{\alpha}_{t} + A_{\alpha}s_{\alpha,t},
   	\end{split}
   \end{equation}
   for $ t = q\times3$, $q = 1,2, \ldots ,T/3$. See~\citet{petrella_2020} for details on the scaled score $s_{\alpha,t}$.
   
\section{Nowcasting US GDP growth in real time}\label{sec:real_time}
	The empirical application is centred on real quarterly GDP, the leading measure of economic activity, and the data are taken from the United States. The six models presented in the previous section are estimated in real-time alongside a Student's t AR(2) model. Each model is estimated recursively every month using an expanding window size mimicking the gradual release of data in real-time. The nowcasting exercise starts in February 2007 and ends with the vintage of April 2023.
			 
	\subsection{Data}
	Historical vintages of the real-time data are taken from Fred-MD, while the Philadelphia Fed provides the vintages of US real GDP. The sample includes data from January 1959 up to April 2023. GDP is modelled in log differences while the Fred-MD data are transformed using the recommended transformations of \citet{mccracken_fred-md_2016}. The data need not be stationary given the presence of idiosyncratic trends in the locations; however, transforming the data is useful to account for level shifts arising from occasional benchmark updates.
	
	Fred-MD contains over a hundred series split into eight groups. Four groups are used in this application: output and income, labor market, consumption order and inventories and stock market. From these four groups, only headline series consistently available from 2007 are used, which leaves ten series : real personal income, industrial production, capacity utilization, civilian employment, total nonfarm employment, civilian unemployment rate, real personal consumption expenditures, retail and food services sales, consumer sentiment index, S\&P's common stock price index: composite. The number of series is constrained by the computing time taken during the real-time nowcasting exercise. The relatively limited number of series used for estimation is not an obstacle to accurate and correctly calibrated predictions, as this section illustrates.
	
	\subsubsection*{Nowcasting steps and the real-time data flow}
	Nowcasts are produced three times between the release of the most recent GDP figure and the next release. For instance, if the calendar quarter of interest ends in month $t$, the first nowcast takes place at the end of month $t-1$ (horizon $h = 2$); real-time data are available for $t-2$ but missing for $t-1$ and $t$. The second nowcasting step occurs at the end of month $t$ when real-time data are available for the first two months of the quarter but missing for the last (horizon $h = 1$). Finally, the third and last nowcast is produced towards the middle or end of $t+1$, before the official release, when real-time data are available for all months in the quarter (horizon $h = 0$).
	
	\subsection{Accounting for the uncertainty arising from missing data}\label{sec:sim_pred}
	At the last nowcasting step, when all the series except GDP are observed, the GDP density nowcast is a direct by-product of estimation. In the mixed-frequency factor model it is given by plugging the filtered location, scale and shape parameters of GDP growth in the skew-t density function. In the factor augmented MIDAS model the predicted parameters are used instead (they are modelled as a function of the estimated filtered common factors). For earlier steps, however, it is important to account for the uncertainty arising from the missing figures in the real-time series; this is done somewhat differently in the mixed-frequency factor model and factor-augmented MIDAS model. See Appendix \ref{ap:uncertainty} for a step by step explanation.
	 
	 \subsection{Evaluating density nowcasts} 
	 
	 Each model's performance is evaluated using two metrics. First, models should provide a reliable measure of nowcasting uncertainty, that is they should be calibrated correctly \citep{mitchell_hall_2005, mitchell_wallis_2011}. Correct calibration can be tested formally using the tests of \citet{diebold_gunther_tay_1998}, \citet{rossi_sekhposyan_2019}, \citet{anderson_darling_1954}, \citet{ljung_box_1978}, and \citet{berkowitz_2001}\footnote{More complicated tests accounting for serial correlation are not needed here because the figure targeted is the latest GDP release only. In this case the nowcasting horizon never exceeds the periodicity of the target variable. Be it a one-month, two-month or three-month ahead, the next nowcast will be generated only after the current observation of GDP has been released. Therefore, the density nowcasts produced here should be independent regardless of their horizon.}. These tests, subsequently referred to as DGT, RS, AD, LB, and Berkowitz tests, exploit the probability integral transforms (PITs) of the outcomes in the nowcast distributions (or their inverse normal transforms for the Berkowitz test). Correctly-calibrated models should exhibit PITs that are uniformally distributed. 
	 
	 Conditional on being correctly calibrated, the best model should yield sharper nowcasts \citep{mitchell_wallis_2011, gneiting_balabdaoui_raftery_2007}. This paper uses two popular sharpness statistics to measure predictive performance: the log score and the continuous rank probability scoring (CRPS) \citep[see][for a detailed discussion of scoring rules]{gneiting_raftery_2007}. The CRPS is less sensitive to large prediction errors than the log score. At the last nowcasting step the predictive distribution is known analytically, as it belongs to the skew-t, but for horizons greater than one the log score and CRPS are evaluated using the empirical distributions computed from the simulated predictions (see section \ref{sec:sim_pred}). The modified version of the \citet{Diebold1995} test from \citet{Harvey1997} is employed to test equal predictive ability.
	 
	 The target variable in this nowcasting exercise is the advance estimate of GDP growth in the US published by the BEA, the timeliest measure of US GDP growth.

	\section{Results}\label{sec:results}
	\subsection{Real-time common factors and their loadings}

\begin{figure}[!h]
		\centering
		\includegraphics[width=0.8\textwidth]{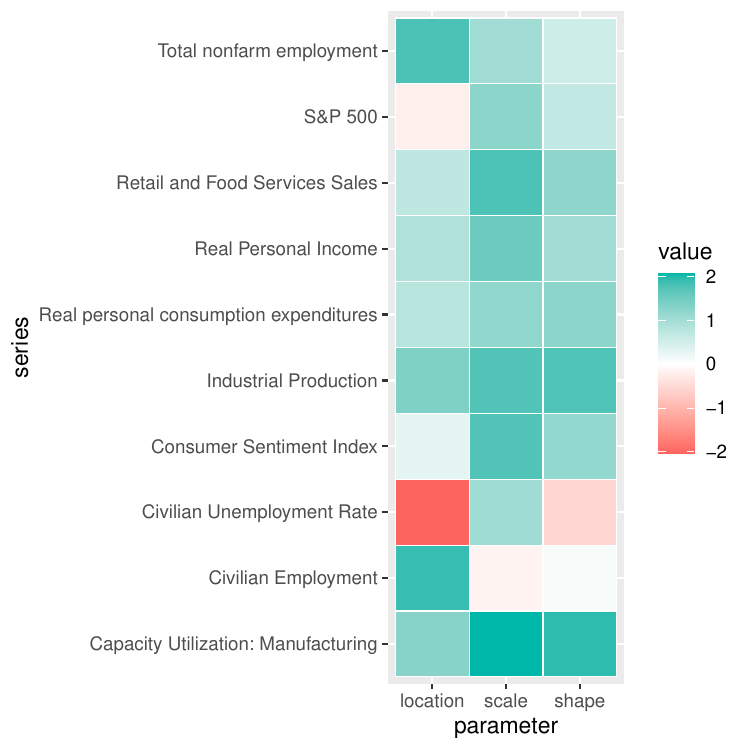}
		\vspace{-10pt}
		\caption{Heat map of the average factor loadings derived from the real-time estimation of the quarterly factor model featuring location, scale and shape common factors.\label{fig:factor_loadings}}
	\end{figure}

	The factor loadings on GDP growth are constrained to one when using the mixed-frequency factor model. In addition to helping for identification, this provides a means for interpreting the impact of each series on the conditional distribution of GDP growth. Figure \ref{fig:factor_loadings} shows the heat map of the factor loadings estimated with the mixed-frequency factor model featuring location, scale and shape common factors. The factor loadings vary during the real-time estimation exercise as the dataset gradually expands; the figure thus shows the average factor loadings attached to the location, scale and shape common factors for each series across the entire real-time exercise. Positive average factor loadings are shown in green and negative average factor loadings in red; the intensity of the colour increases with the absolute value of the average.
	
	First, all series but one exhibit a positive relationship in their conditional dispersion; civilian employment has a negative factor loading on its scale on average but of a very small magnitude. Thus macroeconomic data generally do not become more predictable when general forecasting uncertainty increases. Secondly, the cross-sectional dependencies in locations and shapes mostly go in the same direction; that is, downside risk on a given series is positively related to negative business cycle movements. The S\&P series is the only exception with a negative factor loading in its location on average but a positive factor loading on its shape.
	
	What is more interesting, though, is that the scale and shape common factors capture relationships between series which can be unrelated through their locations. For instance, the S\&P 500 series is only weakly related to the location common component; however, there is a stronger dependence on the scale common component: increased volatility in the stock market is a good predictor of nowcasting uncertainty.
	
	Separately, while survey consumer data cannot provide a clear signal on the location of GDP growth, it is a useful indicator of the shape and magnitude of nowcasting uncertainty. This type of dependence - where series are related through their scales and shapes but not their locations - cannot be captured without the current generation of dynamic factor models because they rely only on common factors in the conditional means. 
	
\begin{figure}[!h]
	\centering
	\includegraphics[width=1\textwidth]{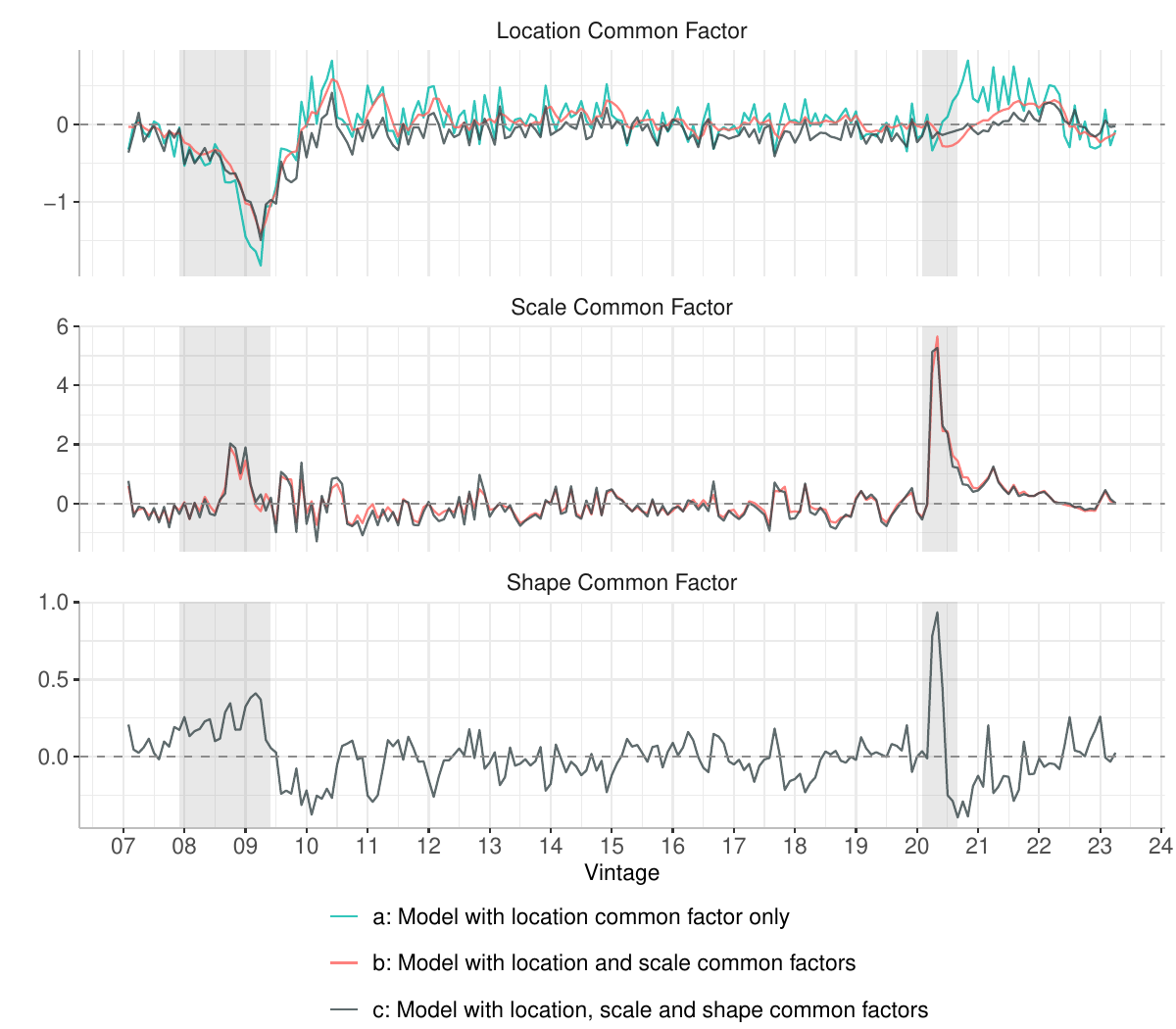}
	\vspace{-10pt}
	\caption{Real-time common factors derived with the different specification of the monthly factor model. These are real-time filtered estimates relating to the last available month in the vintage.\label{fig:real_time_factor_loadings}}
\end{figure}

	Figure \ref{fig:real_time_factor_loadings} shows the real-time path of the location, scale and shape common factors derived with the monthly factor model. In normal times, that is in the expansionary phase of the business cycle, there is not much movement and all factors fluctuate around zero, their long-term average. During the financial crisis the location common factor decreases significantly while the scale common factor increases. This is coherent with the general idea that macroeconomic volatility and the business cycle move counter-cyclically. The Covid-19 induced recession triggers an unprecedented jump in the scale common factor, reflecting unprecedented macroeconomic uncertainty. Overall the scale common factor provides an insight on macroeconomic uncertainty.

	The increase in forecasting uncertainty when entering recessions is generally associated with a sharp increase in the skewness parameter: the probability of negative prediction errors increases while positive prediction errors become less likely. This movement in the skewness parameter is then reversed during recoveries; this is especially apparent during the covid area, where the shape common factor captures the upside risk to economic activity during the early recovery phase. Thus the shape common factor captures both real-time downside and upside risks.
	
	Figure \ref{fig:real_time_factor_loadings} provides one of the first measure of \textit{real-time} macroeconomic asymmetry, or skewness. This in contrast to series-specific (most often GDP) asymmetry, which has been the subject of much attention in economics since the work of \citet{Adrian_2019}. A recent exception in that respect comes from \citet{iseringhausen_petrella_theodoridis_2022} who use the entire Fred-MD database to derive a measure of expected macroeconomic skewness. They first retrieve a measure of conditional skewness for each series independently and then use the first principal component of those as an estimator of aggregate skewness. While their approach has the merit of using a larger number of series than this paper does, the framework presented here has the ability to capture co-movements in skewness across series in real-time. Thus the framework presented in this paper is complementary to theirs. 
	
	Overall the scale and shape common factors carry economic meaning and give a consistent picture of real-time forecasting uncertainty and risk since the financial crisis. As such they can be interpreted respectively as general uncertainty and risk common factors. Next the gain in nowcasting performance stemming from using scale and shape common factors is evaluated.
	
	\subsection{Calibration}
\begin{figure}[H]
	\centering
	\includegraphics[width=1\textwidth]{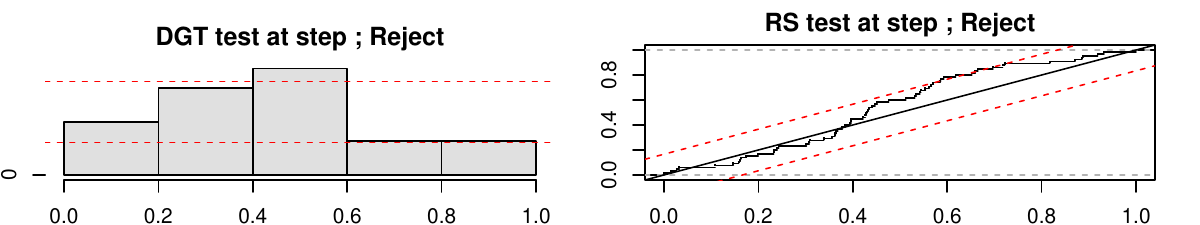}
	\vspace{-20pt}
	\caption{Student's t AR2 model. Graphical representation of the calibration tests of \citet{diebold_gunther_tay_1998} (DGT) and \citet{rossi_sekhposyan_2019} (RS). Null hypothesis of correct calibration. The area between the red dotted line corresponds to the  95\% confidence region.  The number of observations is 61.\label{fig:test_AR2}}
\end{figure}

\begin{figure}[!h]
	\centering
	\includegraphics[width=0.98\textwidth]{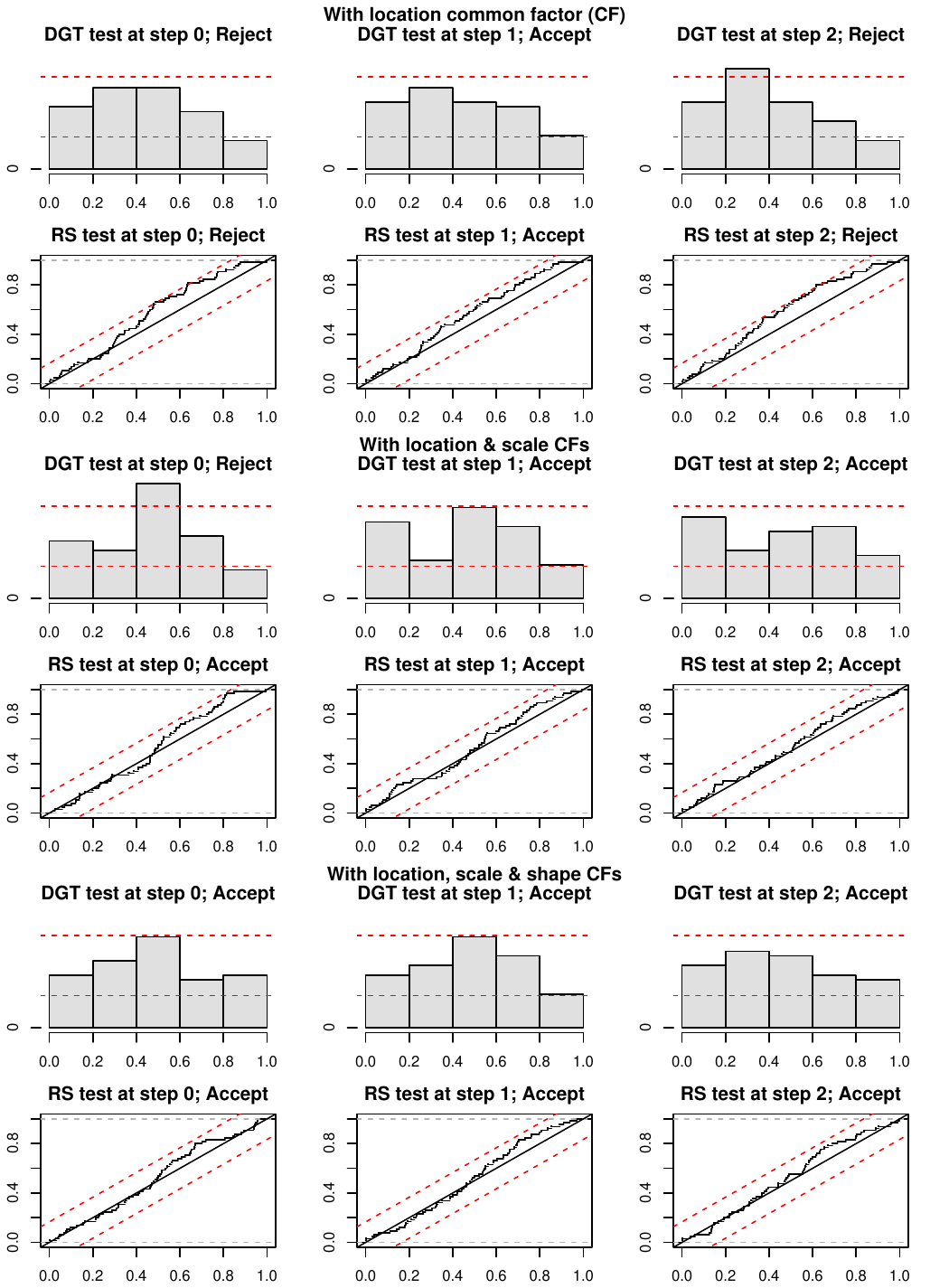}
	\vspace{-0pt}
	\caption{Skew t Mixed-Frequency Factor Model. Graphical representation of the calibration tests of \citet{diebold_gunther_tay_1998} (DGT) and \citet{rossi_sekhposyan_2019} (RS) at horizons $h=0,1,2$. Null hypothesis of correct calibration. The area between the red dotted line corresponds to the  95\% confidence region.  The number of observations is 61 at each step.\label{fig:test_DFM}}
\end{figure}

\begin{figure}[!h]
	\centering
	\includegraphics[width=0.98\textwidth]{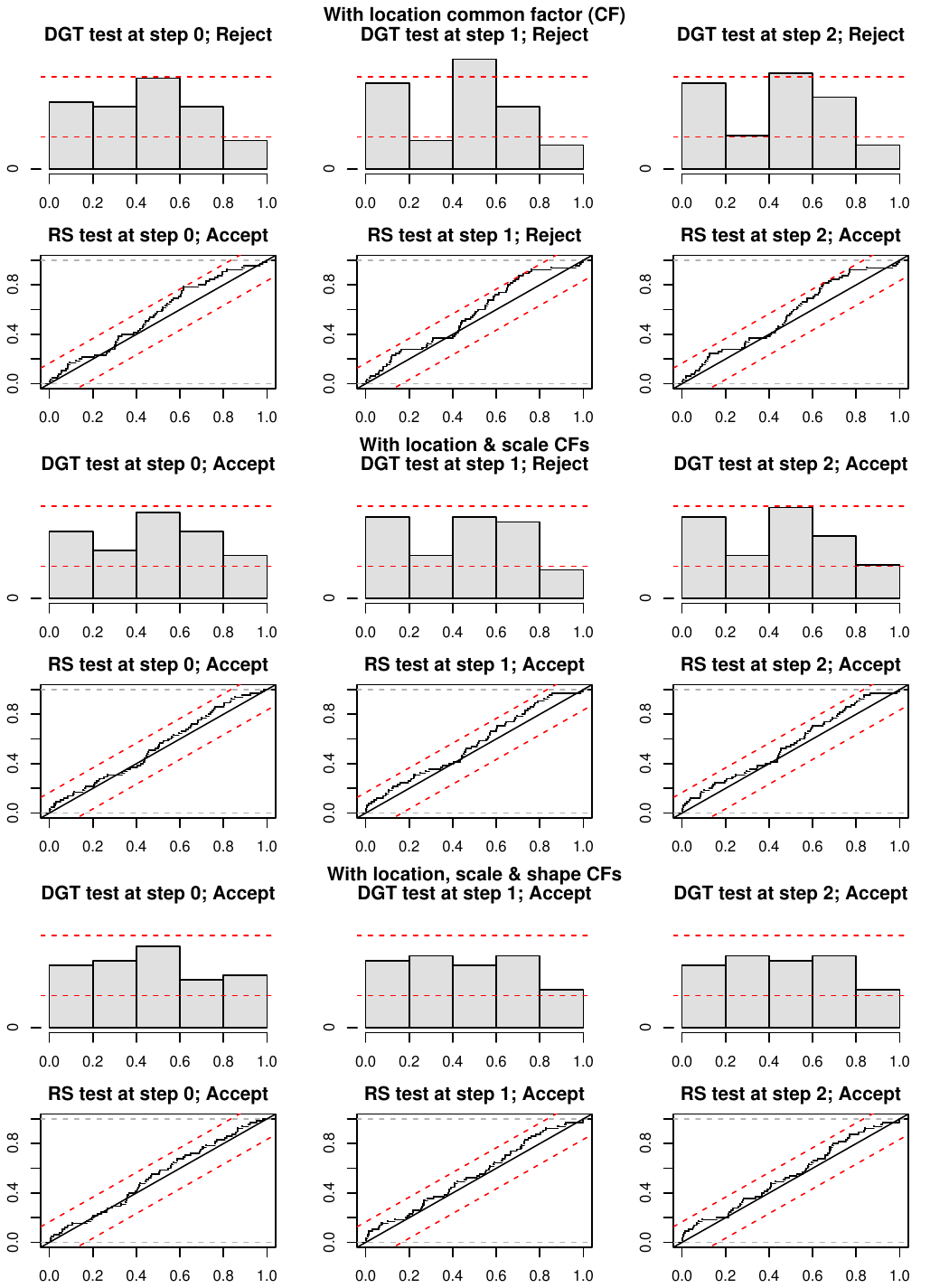}
	\vspace{-0pt}
	\caption{Skew t Factor Augmented MIDAS. Graphical representation of the calibration tests of \citet{diebold_gunther_tay_1998} (DGT) and \citet{rossi_sekhposyan_2019} (RS) at horizons $h=0,1,2$. Null hypothesis of correct calibration. The area between the red dotted line corresponds to the  95\% confidence region.  The number of observations is 61 at each step.\label{fig:test_FA}}
\end{figure}

Table \ref{tb:calibration} shows the statistics of the AD, LB and Berkowitz tests of correct calibration. The graphical representations of the DGT and RS tests are shown in figures \ref{fig:test_AR2}, \ref{fig:test_DFM} and \ref{fig:test_FA}. The DGT test can be subject to false rejections because of its repeated nature and as such is more conservative than the RS test. Note also that the Berkowitz test is highly sensitive to the large prediction errors occurring in 2020 at horizons one and two, which explains its abnormally large values.

\begin{table}[!h]\vspace{5pt} \centering 
	\caption{Statistical tests evaluating out-of-sample calibration at the three nowcasting horizons. The table shows test statistics. AD = Anderson-Darling (Null hypothesis of uniformity and independence of the PITs.); LB = Ljung-Box (Null hypothesis of independence of the PITs (no serial correlation for lags up to four)); Berkowitz = \citet{berkowitz_2001} (Null hypothesis of i.i.d N(0,1) after applying the normal inverse transform to the PITs). ***, ** and * indicate p-values inferior to 0.01, 0.05 and 0.1.\label{tb:calibration}}
	\begin{tabular}{@{\extracolsep{-0pt}} lcccc} 
		\\[-5ex]\hline 
		\hline \\[-1.8ex] 
		Common factors & Horizon & AD & LB & Berkowitz\\
		\hline 
		\hline \\[-1.8ex]
		\multicolumn{5}{c}{Skew t Factor Augmented MIDAS Model} \\ [0.5ex]
Location, scale and shape  & 0 & 0.78 & 6 & 1.26 \\
Location, scale and shape  & 1 & 2.68** & 5 & 37.27*** \\
Location, scale and shape  & 2 & 2.77** & 4.95 & 37.03*** \\
Location, scale & 0 & 1.19 & 5.88 & 4.7 \\
Location, scale & 1 & 2.95** & 5.87 & 30.62*** \\
Location, scale & 2 & 3.08** & 5.42 & 33.5*** \\
Location & 0 & 2.04* & 4.01 & 6.76* \\
Location & 1 & 2.82** & 2.83 & 9.05** \\
Location & 2 & 2.74** & 2.97 & 10.38** \\ 
		\hline 
		\hline \\[-1.8ex]
		\multicolumn{5}{c}{Skew t Mixed-Frequency Factor Model} \\ [0.5ex]
Location, scale and shape & 0 & 0.81 & 7.13 & 0.32 \\
Location, scale and shape & 1 & 1.25 & 8.95* & 1.99 \\
Location, scale and shape & 2 & 1.61 & 2.4 & 13.22*** \\
Location, scale & 0 & 1.84 & 6.5 & 7.09* \\
Location, scale & 1 & 1.22 & 7.48 & 2 \\
Location, scale & 2 & 0.79 & 7.77 & 3.99 \\
Location & 0 & 3.02** & 3.25 & 5.46 \\
Location & 1 & 2.34* & 4.33 & 5.88 \\
Location & 2 & 3.55** & 3.03 & 9.18** \\ 
		\hline 
		\hline \\[-1.8ex]
		\multicolumn{5}{c}{Student's t AR2 Model} \\ [0.5ex]
/ & 1,2,3 & 2.89** & 4.36 & 1.99\\ 
		\hline \\[-1.8ex] 
	\end{tabular} 
\end{table}

The tests for the benchmark AR2 model point clearly against correct calibration. The histogram of the PITs takes a pyramidal shape, indicating that too many observations fall in the centre of the distribution. This occurs when the density nowcasts are too wide; the model overstates dispersion. Overdispersion arises here because the scale parameter is constant and in-sample dispersion is significantly larger than out-of-sample dispersion. Overall the uncertainty attached to the predictions given by the AR2 model are unreliable.

The models exploiting a common factor in the location only also produce poorly calibrated nowcasts with several rejections of both the DGT and RS tests. Introducing a scale common factor improves calibration by removing any rejections of the RS test at 5\%. The improvement in calibration stemming from modelling time-varying volatility has also been pointed out by \citet{Petrella_2020_bis} who model a location common factor featuring stochastic volatility. 

Overall, while it is difficult to avoid rejection in all five calibrations tests, the mixed-frequency dynamic factor model with shape common factor produces the most reliable measure of uncertainty. The PITs do not exhibit the typical pyramidal shape characterising densities which are too wide, and the histogram is rather symmetric, indicating the absence of a bias. Moreover, the RS test is comfortably within the 95\% confidence interval. Modelling a shape common factor also helps improving the calibration of the factor-augmented model.

	\subsection{Sharpness}
\begin{figure}[!h]
		\centering
		\includegraphics[width=0.99\textwidth]{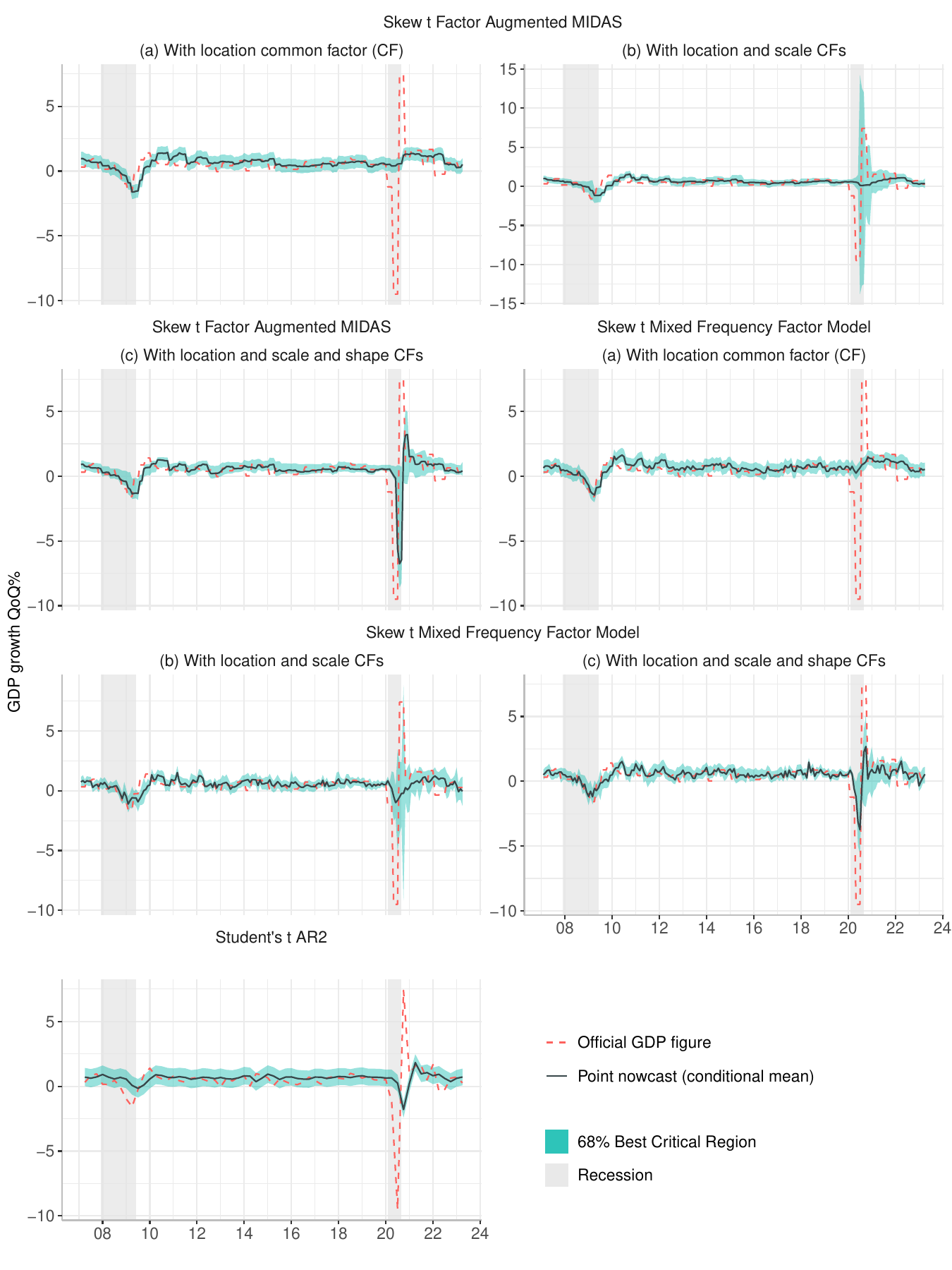}
		\vspace{-0pt}
		\caption{Real-time point nowcast (conditional mean) with the 68\% best critical region for each nowcasting approach and model specification. The x-axis shows the month when the nowcast is produced. The red dotted line shows the first release of GDP (the Advance Estimate) which is the target. The grey-shaded areas indicate the periods of US economic recessions classified by the NBER (The third quarter of 2020 is added for the pandemic).\label{fig:prediction_over_time}}
	\end{figure}

	Figure \ref{fig:prediction_over_time} provides a first insight on the sharpness of each model's nowcasts. It shows the point nowcast for quarterly GDP growth with a 68\% best critical region\footnote{In the presence of skewness or asymmetry, it is more intuitive to use the best critical region instead of a symmetric confidence interval \citep{Mitchell_weale}. The $\alpha$\% BCR represents the shortest region encompassing 68\% probability. The BCR diverges from the confidence interval if the distribution is asymmetric; notably, its frontiers always have equal probabilities. The BCR is equivalent to the Bayesian highest posterior density, used by \citet{Petrella_2020_bis} for illustrating density nowcasts for instance.} (BCR) for each month of the real-time nowcasting exercise. The overdispersion of density nowcasts from models featuring constant scale parameters discussed above is visible in BCRs which are clearly too wide. Separately, the uncertainty around the point nowcast should naturally decrease during the quarter as missing figures in real-time data are gradually replaced with their observations. The AR2 model, however, does not make use of real-time data; its nowcasts are consequently not updated during the quarter. The lack of real-time information in the AR2 model is also particularly apparent in times of recession where the model clearly lag GDP growth. 
	
	Figure \ref{fig:prediction_over_time} is also useful to visualise the unprecedented magnitude of economic fluctuations during the pandemic and understand the unique challenge facing forecasters during this period. Notably, the models using location and scale common factors only are not able to track economic activity and provide a reliable measure of nowcasting uncertainty during this period. While exploiting a scale common factor does generate a large increase in general uncertainty during the pandemic, the lack of movement in the location and the symmetric increase in dispersion means that this uncertainty measure cannot be trusted: there is a significant possibility of large positive figures during the recession and negative figures during the recovery. The models with a shape common factor, on the other hand, are able to track nowcasting uncertainty much more realistically.
	
	\subsubsection*{Sharpness across the nowcasting horizon} 
	Table \ref{tb:relative_perf} shows relative predictive performances across the nowcasting horizon for each model class separately (factor augmented and mixed-frequency factor model) where the models with location common factors are taken as benchmarks. Values less than one indicate a superior performance compared to the location specification. The corresponding test statistics from \citet{Harvey1997} are shown in parenthesis. Overall models with scale and shape common factors tend to do better than models with only a location common factor. This benefit in forecasting performance from modelling scale and shape common factors also increases as more data accrue during the quarter.
	
	Although the models with scale and shape common factor yield sharper nowcasts, these improvements are not generally significant statistically. The factor-augmented model featuring a shape common factor at the last nowcasting step is the only specification rejecting comfortably the test of equal predictive ability. To shade light on this lack of significant improvement in sharpness it is useful to analyse the relative performances of the models across the business cycle.
	
\begin{table}[!htbp] \centering \caption{Relative predictive performance across the nowcasting horizon as measured with the (negative) log score and CRPS. The table reports the performance of models with scale and shape factors, as well as the AR2, compared with the model with location common factor only, which is used as denominator; hence a ratio lower than one means that the model with location common factor only is outperformed. The corresponding statistics from the modified Diebolt and Mariano test of \citet{Harvey1997} is shown in parenthesis. ***, ** and * indicate p-values inferior to 0.01, 0.05 and 0.1. h = 0 refers to the last nowcasting step before the release of the advance GDP estimate. FA = Factor-augmented; LS = Location and Scale; LSS = Location scale and shape; MF-FM = Mixed-frequency factor model}   \label{tb:relative_perf} \begin{tabular}{@{\extracolsep{-5pt}} lcccccccc} \\[-1.8ex]\hline
	&	\multicolumn{4}{c}{CRPS} & 	\multicolumn{4}{c}{Log Score}\\
		 \hline \\[-1.8ex] Model & h = 0 & h = 1 & h = 2 & mean & h = 0 & h = 1 & h = 2 & mean \\ 		\hline 
		 \hline \\[-1.8ex]
		\multicolumn{9}{c}{Performance relative to the factor-augmented MIDAS model with location factor} \\ [0.5ex]
FA LS & 0.874 & 0.98 & 0.986 & 0.948 & 0.798 & 0.96 & 1.073 & 0.946 \\ 
& (-1.061) & (-0.287) & (-0.209) &  & (-1.118) & (-0.2) & (0.486) &  \\ \\[-1.8ex]
FA LSS & 0.668 & 1.051 & 1.064 & 0.933 & 0.566*** & 0.903 & 1.005 & 0.829 \\ 
& (-1.599)& (0.4) & (0.497) &  & (-2.466) & (-0.74) & (0.053) &  \\ \\[-1.8ex]
Student's t AR2 & 1.113 & 1.052* & 1.048 & 1.07 & 1.044 & 0.935 & 1.013 & 0.995 \\ 
& (1.586) & (1.428) & (1.398) &  & (1.398) & (-0.189) & (1.649) &  \\
		\hline 
\hline \\[-1.8ex]
\multicolumn{9}{c}{Performance relative to the mixed-frequency factor model with location factor} \\ [0.5ex]
MF-FM LS & 0.821 & 0.973 & 1.001 & 0.933 & 0.648* & 0.743 & 0.985 & 0.79 \\ 
& (-1.476) & (-0.392) & (0.009) &  & (-1.904) & (-1.228) & (-0.089) &  \\ \\[-1.8ex]
MF-FM LSS & 0.748 & 0.747 & 0.98 & 0.826 & 0.76 & 0.66 & 1.079 & 0.825 \\
 & (-1.268) & (-1.569) & (-0.291) &  & (-1.159) & (-1.346) & (0.631) &  \\ \\[-1.8ex]
Student's t AR2 & 1.163 & 1.137 & 1.126 & 1.142 & 1.124 & 0.97 & 1.109 & 1.063 \\
  & (1.075) & (0.612) & (0.556) &  & (0.469) & (-0.545) & (0.146) &  \\ \hline \\[-1.8ex] \end{tabular} \end{table} 
	
	\subsubsection*{Sharpness across the business cycle}
\begin{figure}[!h]
		\centering
		\includegraphics[width=1\textwidth]{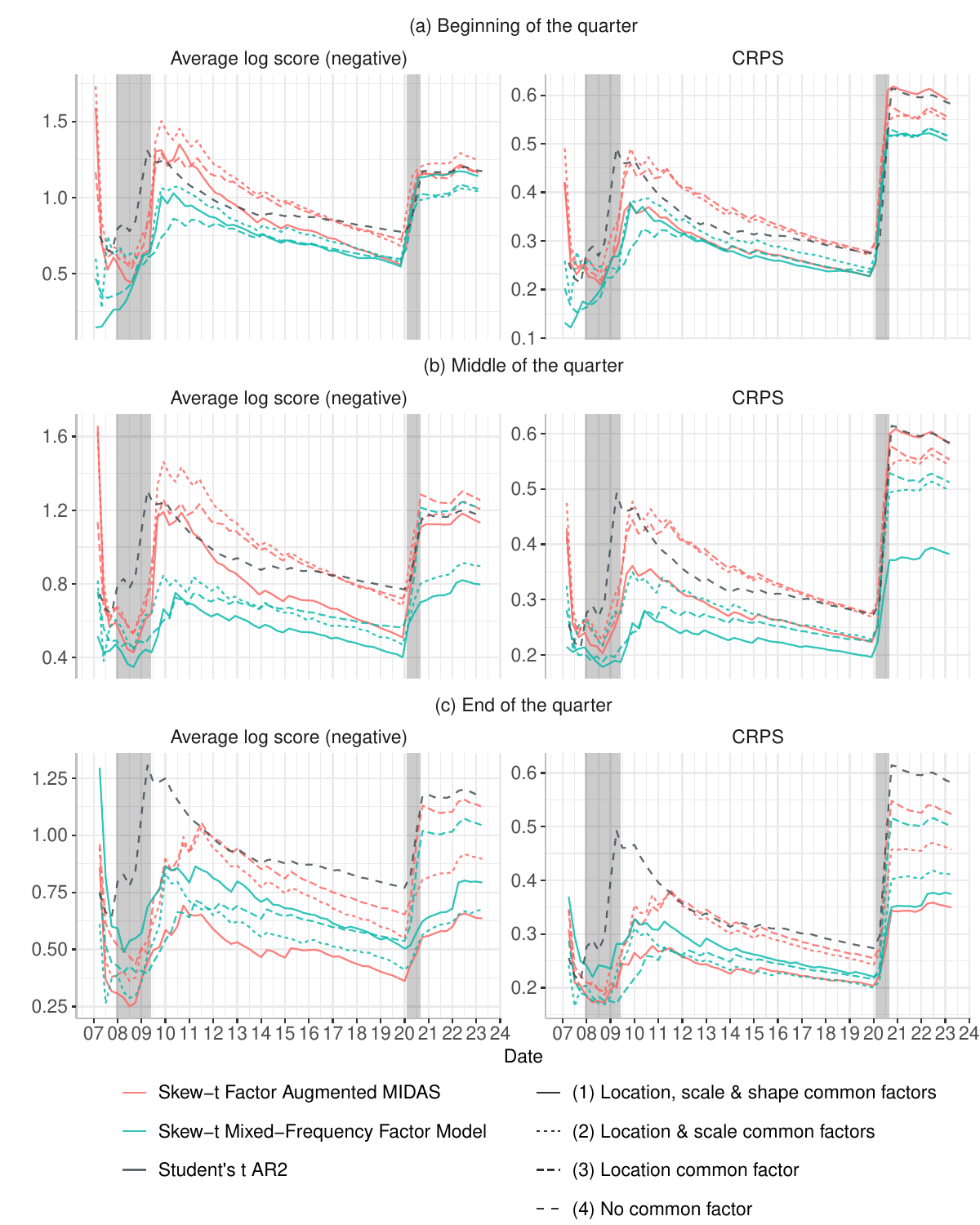}
		\vspace{-10pt}
		\caption{Average statistics over time (cumulated statistics) derived from real-time estimation. The sample used to compute the statistics grows with the expanding estimation window. The grey-shaded areas indicate the periods of US economic recessions classified by the NBER (The third quarter of 2020 is added for the pandemic).\label{fig:cummulated_stats}}
	\end{figure}

	Table \ref{tb:sharpness_stats} shows the average log score and CRPS during the recession of 2007 financial crisis 2007 Q4 - 2009 Q2, the Covid-19 pandemic (2020 Q1 -  2020 Q3) and remaining periods. The third quarter of 2020 is included because it featured an unusually large increase in economic activity; forecasting such movements matters for policymakers given that excessively large growth generates price pressures. Figure \ref{fig:cummulated_stats} shows the average of the statistics across the real-time nowcasting exercise (cumulated average); the last point of these lines thus coincides with the overall average.
	
	Exploiting scale and shape common factors does not materially improve sharpness during the long expansionary phase of the 2010s. All models including the AR2 do relatively well and their ranking is steady during the 2010s. Modelling a location common factor only also yields good results for the financial crisis when using a mixed-frequency factor specification. But note that although scale and shape common factors do not yield sharper nowcasts in this case, they remain important to produce reliable measures of uncertainty; as discussed above, the mixed-frequency factor model with only a location common factor is poorly calibrated. 

\begin{table}[!htbp] \centering 
	\caption{Real-time statistics examining the sharpness of density nowcasts. Bold values indicate the sharpest nowcast for each period and statistic. Expansion refers to normal times. GFC = Great Financial Crisis. The Covid period includes both the pandemic recession and subsequent recovery quarter. ALS = Average log score (the higher the better).} 
	\label{tb:sharpness_stats} 
	\begin{tabular}{@{\extracolsep{20pt}} llcc} 
		\\[-4ex]\hline 
		\hline \\[-1.8ex]
		Period & Common factor &CRPS & ALS\\
		\hline 
		\hline \\[-1.8ex]
		\multicolumn{4}{c}{Skew t Factor Augmented MIDAS Model} \\ [0.5ex]
2007 Q1 - 2023 Q1 & Location & 0.28 & -0.73 \\
except Covid-19 and GFC & Location, scale & 0.3 & -0.77 \\
 & Location, scale, shape & \textbf{0.25} & \textbf{-0.59} \\  \hdashline[3pt/5pt] \\ [-2.5ex]
 & Location & 0.35 & -0.96 \\
 (GFC)& Location, scale & 0.35 & -0.83 \\
 & Location, scale, shape & 0.3 & -0.65 \\  \hdashline[3pt/5pt] \\ [-2.5ex]
 & Location & 5.77 & -9.84 \\
2020 Q1 -  2020 Q3 (Covid-19)& Location, scale & 4.89 & -8.14 \\
 & Location, scale, shape & 5.71 & -8.89 \\
		\hline 
		\hline \\[-1.8ex]
		\multicolumn{4}{c}{Skew t Mixed-Frequency Factor Model} \\ [0.5ex]  
2007 Q1 - 2023 Q1 & Location & 0.26 & -0.68 \\
except Covid-19 and GFC & Location, scale & 0.27 & -0.63 \\
 & Location, scale, shape & \textbf{0.25} & -0.63 \\  \hdashline[3pt/5pt] \\ [-2.5ex]
 & Location & \textbf{0.23} & \textbf{-0.57} \\
2007 Q4 - 2009 Q2 (GFC)  & Location, scale & 0.26 & -0.61 \\
 & Location, scale, shape & 0.26 & -0.68\\  \hdashline[3pt/5pt] \\ [-2.5ex]
 & Location & 5.72 & -10.11 \\
2020 Q1 -  2020 Q3 (Covid-19)& Location, scale & 4.76 & \textbf{-5.95} \\
 & Location, scale, shape & \textbf{3.85} & -6.63 \\ 
		\hline 
		\hline \\[-1.8ex]
		\multicolumn{4}{c}{Student's t AR2 Model} \\ [0.5ex]
2007 Q1 - 2023 Q1 & No common factor & 0.26 & -0.75 \\
except Covid-19 and GFC& & &  \\
2007 Q4 - 2009 Q2 (GFC) & No common factor & 0.57 & -1.48 \\
2020 Q1 -  2020 Q3 (Covid-19) & No common factor & 6.52 & -8.14 \\ 
		\hline \\[-1.8ex] 
	\end{tabular} 
\end{table} 
	
\begin{figure}[!h]
		\centering
		\includegraphics[width=1\textwidth]{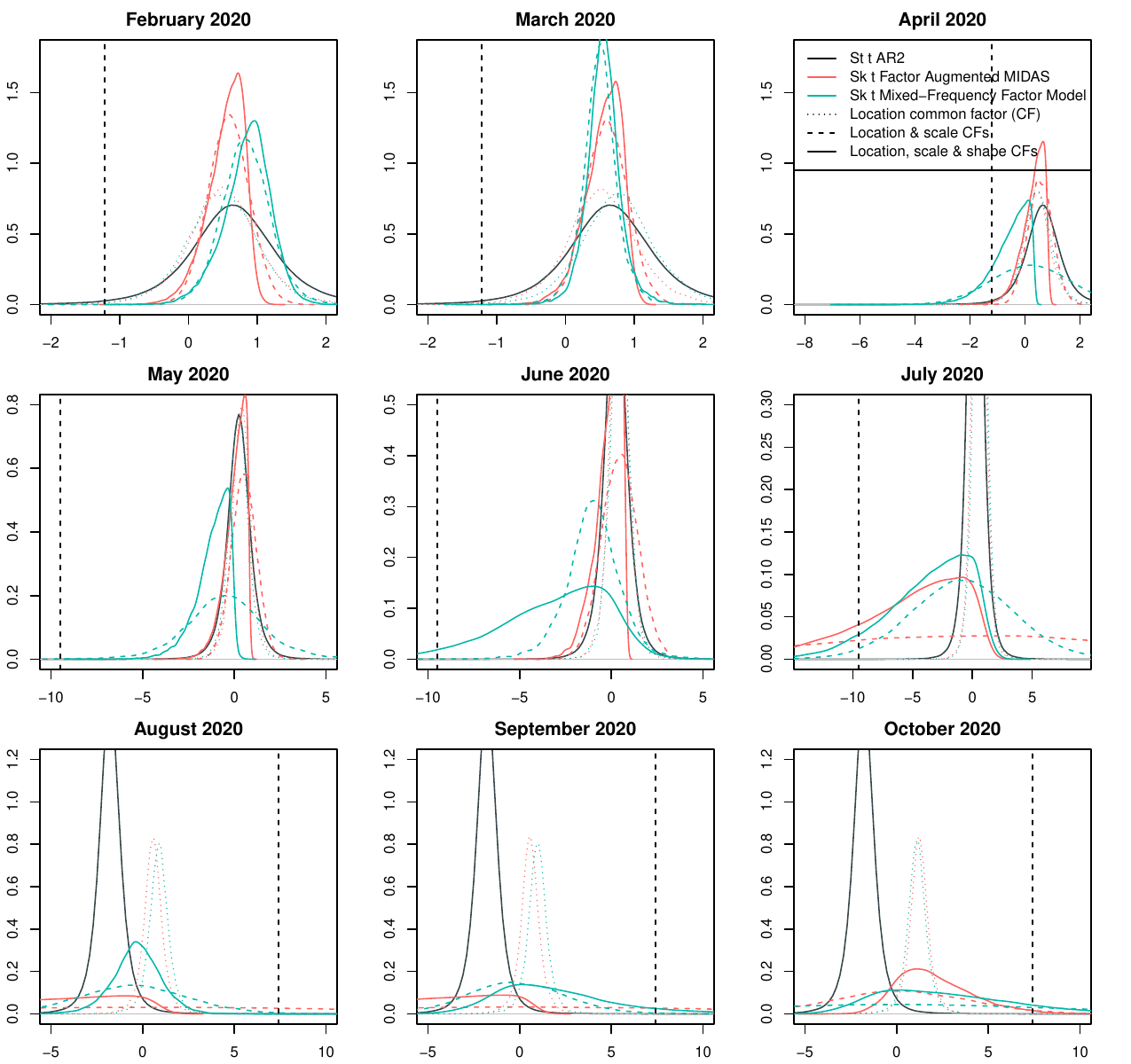}
		\vspace{-20pt}
		\caption{Illustrating the importance of the shape common factor during the pandemic. Real-time density nowcasts during the early phase of the pandemic. Vertical lines indicate the first release of GDP growth (the BEA Advance Estimate).\label{fig:covid_densities}}
	\end{figure}

	The shape common factor proves to be most useful to capture the unprecedented economic fluctuations occurring during the first three quarters of 2020. Figure \ref{fig:covid_densities} shows density nowcasts at each nowcasting step from February to October 2020. Focusing on the first quarter, the effect of the government restrictions became visible only in the March data published in April. Here the mixed-frequency factor model with shape common factor is the only model able to capture the turning point in economic activity in a timely manner. The density nowcast in April is clearly skewed towards negative prediction errors and reflects the general forecasting environment at the time;  while it was clear that the US economy shrank in the first quarter of 2020, the magnitude of the downturn was uncertain.
	
	While modelling a scale common factor translates into an increased dispersion of the nowcasts, the uncertainty is symmetric and the likelihood of large positive prediction errors increases as well, which is unrealistic. A sharp drop in the location would offset this behaviour, but the data early during the pandemic were not informative enough regarding the location of the density nowcasts. Importantly, the relatively subdued movements in the locations are not a feature of the models studied here; a look at the historical forecasts of the New York Fed and Atlanta Fed\footnote{In mid April 2020, which here corresponds to the last nowcasting step for the first quarter of 2020, GDPNow showed a -0.3\% annualised quarterly growth rate, which corresponds to -0.075\% quarterly growth rate; the New York Fed model showed a -0.39\% annualised quarterly growth rate, which corresponds to about -0.1\% quarterly growth rate. The BEA GDP quarterly growth rate advance estimate for 2021 Q1 was -1.2\%.} (GDPNow) shows that these models did not capture the drop in activity early during the pandemic, which highlights the need to go beyond information in the conditional means of real-time data.
	
	The shape common factor is not only useful to capture the exceptional fall in activity during the pandemic, it also enables both the mixed-frequency factor model and the factor-augmented model to track the sharp rebound in activity that followed. Overall modelling a shape common factor is critical to producing a robust nowcasting performance during the pandemic. The pandemic, and specifically late in the nowcasting window, is the only period where modelling scale and shape common factors materially improve sharpness. This helps explain why the tests of equal predictive ability are generally not rejected.
	
	\section{Conclusion}\label{seq:conclusion}
	
	This paper shows that more information can be extracted from real-time data, namely their conditional dispersion and asymmetry, which improves density nowcasts of GDP growth. For this a skew-t score driven model is derived where the dynamic factor structure commonly applied to location parameters is extended to scale and shape parameters. A real-time estimation exercise shows that scale and shape common factors track general forecasting uncertainty and macroeconomic risk. Further, it is shown how these two novel common factors may be exploited within two widely-used frameworks for nowcasting: mixed-frequency factor models and factor-augmented MIDAS models. Exploiting scale and shape common factors for nowcasting improves calibration and thus helps produce more reliable measures of nowcasting uncertainty. The shape common factor also contributes to sharper predictions by capturing in real-time the asymmetric uncertainty arising during the early phase of the pandemic and subsequent recovery.
	
			{\onehalfspacing
		\bibliography{main}}
	
	\begin{appendices}
		
	\section{Matrix representation of the score driven factor model}\label{ap:score}
	This appendix sets out in full details the transition system \eqref{eq:updating_step}:
	\begin{equation*}\begin{split}
			f_{t+1} &= Bf_{*, t},\\
			f_{*, t} &= f_{t} + Ds_t(f_t).
	\end{split}\end{equation*}

	Sub-models for location, scale and shape parameters can be set out separately and run in parallel during the recursion, making presentation and implementation easier.
	
	\subsection{Location sub-model}
	The transition system for the location component can be written as
			\begin{equation} \begin{split}
			\begin{pmatrix} \bar{\mu}_{1,t+1}\\
				\bar{\mu}_{2,t+1}\\
				\vdots\\
				\bar{\mu}_{N,t+1}\\
				\tilde{\mu}_{t+1}\\
				\tilde{\mu}_{t}\end{pmatrix} & = 
			\begin{pmatrix}1 & 0 & \cdots & 0 & 0 & 0\\
				0 &  1 & \cdots & 0 & 0 & 0\\
				\vdots  & \vdots & \ddots &\vdots & \vdots& \vdots\\
				0 & 0 & \cdots & 1 & 0 &0\\
				0 & 0 & \cdots & 0 & \phi_1 & \phi_2\\
				0 & 0 & \cdots & 0 & 1 & 0\\\end{pmatrix}f^{\mu}_{t|t},\\
				f^{\mu}_{t|t} &=
				\begin{pmatrix} \bar{\mu}_{1,t}\\
					\bar{\mu}_{2,t}\\
					\vdots\\
					\bar{\mu}_{N,t}\\
					\tilde{\mu}_{t}\\
					\tilde{\mu}_{t-1}\end{pmatrix} +
				\begin{pmatrix}a_{\mu,1} & 0 & \cdots & 0 & 0 & 0\\
					0 &  a_{\mu,2} & \cdots & 0 & 0 & 0\\
					\vdots  & \vdots & \ddots &\vdots & \vdots& \vdots\\
					0 & 0 & \cdots & a_{\mu,N} & 0 &0\\
					0 & 0 & \cdots & 0 & a_{\mu,N+1} & 0\\
					0 & 0 & \cdots & 0 & 0 & 0\\\end{pmatrix}
				\begin{pmatrix}  \frac{\partial \text{log}p(y_t|f_t, \Theta)}{\partial \bar{\mu}_{1,t}}\\
					\frac{\partial \text{log}p(y_t|f_t, \Theta)}{\partial \bar{\mu}_{2,t}}\\
					\vdots\\
					\frac{\partial \text{log}p(y_t|f_t, \Theta)}{\partial \bar{\mu}_{N,t}}\\
					\frac{\partial \text{log}p(y_t|f_t, \Theta)}{\partial \tilde{\mu}_{t}}\\
					0\end{pmatrix}.
			\end{split}	\end{equation}
			Since the log density \eqref{eq:log_density_indep} relies on conditional independence, the score can be conveniently written as
\begin{equation}
	\frac{\partial \text{log}p(y_t|f_t, \Theta)}{\partial x} = \sum^N_{i=1}\delta_{i,t}\frac{\partial\text{log}f(y_{i,t}|f_t, \Theta)}{\partial x}.
\end{equation}
The score related to the idiosyncratic component of series i is then
\begin{equation}\begin{split}
	\frac{\partial \text{log}p(y_t|f_t, \Theta)}{\partial \bar{\mu}_{i,t}} &= \sum^N_{j=1}\delta_{i,t}\frac{\partial\text{log}f(y_{i,t}|f_t, \Theta)}{\partial \bar{\mu}_{i,t}},\\
		 &= \delta_{i,t}\frac{\partial\text{log}f(y_{i,t}|f_t, \Theta)}{\partial \bar{\mu}_{i,t}},\\
		 &= \delta_{i,t}\frac{\partial\text{log}f(y_{i,t}|f_t, \Theta)}{\partial \mu_{i,t}}.\frac{\partial\mu_{i,t}}{\partial \bar{\mu}_{i,t}},\\
		 &= \delta_{i,t}\frac{\partial\text{log}f(y_{i,t}|f_t, \Theta)}{\partial \mu_{i,t}},\\
\end{split}\end{equation}
and the score related to the location common factor is
\begin{equation}\begin{split}
		\frac{\partial \text{log}p(y_t|f_t, \Theta)}{\partial \tilde{\mu}_{t}} &= \sum^N_{i=1}\delta_{i,t}\frac{\partial\text{log}f(y_{i,t}|f_t, \Theta)}{\partial \tilde{\mu}_{t}},\\
		&= \sum^N_{i=1}\delta_{i,t}\frac{\partial\text{log}f(y_{i,t}|f_t, \Theta)}{\partial \mu_{i,t}}.\frac{\partial\mu_{i,t}}{\partial \tilde{\mu}_{t}},\\
		&= \sum^N_{i=1}\delta_{i,t}\frac{\partial\text{log}f(y_{i,t}|f_t, \Theta)}{\partial \mu_{i,t}}.\Lambda_{\mu,i},\\
\end{split}\end{equation}
where
\begin{equation}\begin{split}
		\frac{\partial\text{log}f(y_{i,t}|f_t, \Theta)}{\partial \mu_{i,t}} = \frac{1}{\sigma_{i,t}}w_{i,t} \zeta_{i,t},\\
\end{split}\end{equation}
with $w_{i,t} = \frac{(1+\eta_i)}{(1-sgn(y_t-\mu_{i,t})\alpha_{i,t})^2 +\eta_i\zeta^2_{i,t}}$, and $\zeta_{i,t} =  \frac{y_t-\mu_{i,t}}{\sigma_{i,t}}$; see \citet{petrella_2020} for a detailed derivation.
\subsection{Scale sub-model}
The transition system for the scale component can be written as
			\begin{equation*}\begin{split}
			\begin{pmatrix} \bar{\sigma}_{1,t+1}\\
				\bar{\sigma}_{2,t+1}\\
				\vdots\\
				\bar{\sigma}_{N,t+1}\\
				\tilde{\sigma}_{t+1}\end{pmatrix} &= 
			\begin{pmatrix}1 & 0 & \cdots & 0 & 0 \\
				0 &  1 & \cdots & 0 & 0 \\
				\vdots  & \vdots & \ddots &\vdots & \vdots\\
				0 & 0 & \cdots & 1 & 0 \\
				0 & 0 & \cdots & 0 & \phi_1\\\end{pmatrix} f^{\sigma}_{t|t},\\
				f^{\sigma}_{t|t} &= 
			\begin{pmatrix} \bar{\sigma}_{1,t}\\
				\bar{\sigma}_{2,t}\\
				\vdots\\
				\bar{\sigma}_{N,t}\\
				\tilde{\sigma}_{t}\\\end{pmatrix} +
			\begin{pmatrix} a_{\sigma,1}  & 0 & \cdots & 0 & 0 \\
				0 &  a_{\sigma,2} & \cdots & 0 & 0 \\
				\vdots  & \vdots & \ddots &\vdots & \vdots\\
				0 & 0 & \cdots & a_{\sigma,N} & 0\\
				0 & 0 & \cdots & 0 & a_{\sigma,N+1} \\\end{pmatrix}
			\begin{pmatrix}  \frac{\partial \text{log}p(y_t|f_t, \Theta)}{\partial \bar{\sigma}_{1,t}}\\
				\frac{\partial \text{log}p(y_t|f_t, \Theta)}{\partial \bar{\sigma}_{2,t}}\\
				\vdots\\
				\frac{\partial \text{log}p(y_t|f_t, \Theta)}{\partial \bar{\sigma}_{N,t}}\\
				\frac{\partial \text{log}p(y_t|f_t, \Theta)}{\partial \tilde{\sigma}_{t}}\end{pmatrix}.
			\end{split}\end{equation*}
The score related to the idiosyncratic scale component of series i is 
			\begin{equation}\begin{split}
					\frac{\partial \text{log}p(y_t|f_t, \Theta)}{\partial \bar{\sigma}_{i,t}} &= \sum^N_{j=1}\delta_{j,t}\frac{\partial\text{log}f(y_{j,t}|f_t, \Theta)}{\partial \bar{\sigma}_{i,t}},\\
					&= \delta_{i,t}\frac{\partial\text{log}f(y_{i,t}|f_t, \Theta)}{\partial \bar{\sigma}_{i,t}},\\
					&= \delta_{i,t}\frac{\partial\text{log}f(y_{i,t}|f_t, \Theta)}{\partial \sigma_{i,t}}.\frac{\partial\sigma_{i,t}}{\partial \bar{\sigma}_{i,t}},\\
					&= \delta_{i,t}\frac{\partial\text{log}f(y_{i,t}|f_t, \Theta)}{\partial \sigma_{i,t}}.\sigma_{i,t},\\
			\end{split}\end{equation}
and the score related to the scale common factor is			
			\begin{equation}\begin{split}
					\frac{\partial \text{log}p(y_t|f_t, \Theta)}{\partial \tilde{\sigma}_{t}} &= \sum^N_{i=1}\delta_{i,t}\frac{\partial\text{log}f(y_{i,t}|f_t, \Theta)}{\partial \tilde{\sigma}_{t}},\\
					&= \sum^N_{i=1}\delta_{i,t}\frac{\partial\text{log}f(y_{i,t}|f_t, \Theta)}{\partial \sigma_{i,t}}.\frac{\partial\sigma_{i,t}}{\partial \tilde{\sigma}_{t}},\\
					&= \sum^N_{i=1}\delta_{i,t}\frac{\partial\text{log}f(y_{i,t}|f_t, \Theta)}{\partial \sigma_{i,t}}\Lambda_{\sigma,i}\sigma_{i,t},\\
			\end{split}\end{equation}

\begin{equation}\begin{split}
		\frac{\partial\text{log}f(y_{i,t}|f_t, \Theta)}{\partial \sigma_{i,t}} = \frac{1}{\sigma_{i,t}}(w_{i,t}\zeta^2_{i,t}-1).\\
\end{split}\end{equation}
See \citet{petrella_2020} for a detailed derivation.

\subsection{Shape sub-model}
The transition system for the shape component can be written as
			\begin{equation}\begin{split}
			\begin{pmatrix} \bar{\alpha}_{1,t+1}\\
				\bar{\alpha}_{2,t+1}\\
				\vdots\\
				\bar{\alpha}_{N,t+1}\\
				\tilde{\alpha}_{t+1}\end{pmatrix} &= 
			\begin{pmatrix}1 & 0 & \cdots & 0 & 0 \\
				0 &  1 & \cdots & 0 & 0 \\
				\vdots  & \vdots & \ddots &\vdots & \vdots\\
				0 & 0 & \cdots & 1 & 0 \\
				0 & 0 & \cdots & 0 & \phi_1\\\end{pmatrix} f^{\alpha}_{t|t}, \\
			f^{\alpha}_{t|t} & = \begin{pmatrix} \bar{\alpha}_{1,t}\\
				\bar{\alpha}_{2,t}\\
				\vdots\\
				\bar{\alpha}_{N,t}\\
				\tilde{\alpha}_{t}\\\end{pmatrix} +
			\begin{pmatrix}a_{\alpha,1} & 0 & \cdots & 0 & 0 \\
				0 &  a_{\alpha,2}  & \cdots & 0 & 0 \\
				\vdots  & \vdots & \ddots &\vdots & \vdots\\
				0 & 0 & \cdots &a_{\alpha,N}  & 0\\
				0 & 0 & \cdots & 0 & a_{\alpha,N+1}  \\\end{pmatrix}
			\begin{pmatrix}  \frac{\partial \text{log}p(y_t|f_t, \Theta)}{\partial \bar{\alpha}_{1,t}}\\
				\frac{\partial \text{log}p(y_t|f_t, \Theta)}{\partial \bar{\alpha}_{2,t}}\\
				\vdots\\
				\frac{\partial \text{log}p(y_t|f_t, \Theta)}{\partial \bar{\alpha}_{N,t}}\\
				\frac{\partial \text{log}p(y_t|f_t, \Theta)}{\partial \tilde{\alpha}_{t}}\end{pmatrix}.
		\end{split}\end{equation}
The score related to the idiosyncratic shape component of series i is 	
				\begin{equation}\begin{split}
			\frac{\partial \text{log}p(y_t|f_t, \Theta)}{\partial \bar{\alpha}_{i,t}} &= \sum^N_{i=1}\delta_{i,t}\frac{\partial\text{log}f(y_{i,t}|f_t, \Theta)}{\partial \bar{\alpha}_{i,t}},\\
			&= \delta_{i,t}\frac{\partial\text{log}f(y_{i,t}|f_t, \Theta)}{\partial \bar{\alpha}_{i,t}},\\
			&= \delta_{i,t}\frac{\partial\text{log}f(y_{i,t}|f_t, \Theta)}{\partial \alpha_{i,t}}.\frac{\partial\alpha_{i,t}}{\partial \bar{\alpha}_{i,t}},\\
			&= \delta_{i,t}\frac{\partial\text{log}f(y_{i,t}|f_t, \Theta)}{\partial \alpha_{i,t}}.(1-\alpha^2_{i,t}),\\
	\end{split}\end{equation}
The score related to the shape component common factor of series is 		
	\begin{equation}\begin{split}
			\frac{\partial \text{log}p(y_t|f_t, \Theta)}{\partial \tilde{\alpha}_{t}} &= \sum^N_{i=1}\delta_{i,t}\frac{\partial\text{log}f(y_{i,t}|f_t, \Theta)}{\partial \tilde{\alpha}_{t}},\\
			&= \sum^N_{i=1}\delta_{i,t}\frac{\partial\text{log}f(y_{i,t}|f_t, \Theta)}{\partial \alpha_{i,t}}.\frac{\partial\alpha_{i,t}}{\partial \tilde{\alpha}_{t}},\\
			&= \sum^N_{i=1}\delta_{i,t}\frac{\partial\text{log}f(y_{i,t}|f_t, \Theta)}{\partial \alpha_{i,t}}\Lambda_{\alpha,i}(1-\alpha^2_{i,t}),\\
	\end{split}\end{equation}
with
\begin{equation}\begin{split}
		\frac{\partial\text{log}f(y_{i,t}|f_t, \Theta)}{\partial \alpha_{i,t}} = -\frac{sgn(y_{i,t}-\mu_{i,t})}{1-sgn(y_{i,t}-\mu_{i,t})\alpha_{i,t}}(w_{i,t}\zeta^2_{i,t}).\\
\end{split}\end{equation}
See \citet{petrella_2020} for a detailed derivation.

\subsection{Rolling quarterly model}
	The mixed-frequency models featuring scale and shape common factors rely on the data being aggregated into rolling quarterly figures. To accommodate this feature the location sub-model has to be augmented with lags of the unobserved components; precisely the trends and common factor. The scale and shape models remain unchanged. 
	
	Modelling lags in transition equations that take a matrix form is commonly done as follows: 
				\begin{equation}
		\begin{pmatrix}\bar{\mu}_{i,t+1}\\
			\bar{\mu}_{i,t}\\
		\bar{\mu}_{i,t-1}\\
			\bar{\mu}_{i,t-2}\\
			\bar{\mu}_{i,t-3}\end{pmatrix} = 
		\begin{pmatrix}1 & 0 & 0 & 0 & 0 \\
			1 &  0 & 0 & 0 & 0 \\
			0 &  1 & 0 & 0 & 0 \\
			0 &  0 & 1 & 0 & 0 \\
		0 &  0 & 0 & 1 & 0 \\\end{pmatrix}
		\begin{pmatrix} \bar{\mu}_{i,t}\\
		\bar{\mu}_{i,t-1}\\
			\bar{\mu}_{i,t-2}\\
			\bar{\mu}_{i,t-3}\\
		\bar{\mu}_{i,t-4}\end{pmatrix} +
		\begin{pmatrix}a_{\mu,i} & 0 & 0 & 0 & 0\\
			0 & 0 & 0 & 0 & 0\\
			0 & 0 & 0 & 0 & 0\\
			0 & 0 & 0 & 0 & 0\\
			0 & 0 & 0 & 0 & 0\end{pmatrix}
		\begin{pmatrix}  \frac{\partial \text{log}p(y_t|f_t, \Theta)}{\partial \bar{\mu}_{i,t}}\\
			0\\
			0\\
			0\\
			0\end{pmatrix},
	\end{equation}
for $i = 1,...,N$. Augmenting the common factor with lags is done in exactly the same way. The equation linking the monthly components to the rolling quarterly series $i$ becomes: 
 \begin{equation}\begin{split}\mu_{i,t} &= \frac{1}{3}\bar{\mu}_{i,t}+\frac{2}{3}\bar{\mu}_{i,t-1}+ \bar{\mu}_{i,t-2} + \frac{2}{3}\bar{\mu}_{i,t-3} +\frac{1}{3}\bar{\mu}_{i,t-4} \\&+ \Lambda_{\mu_i}(\frac{1}{3}\tilde{\mu}_{i,t}+\frac{2}{3}\tilde{\mu}_{i,t-1}+\tilde{\mu}_{i,t-2} +\frac{2}{3}\tilde{\mu}_{i,t-3} +\frac{1}{3}\tilde{\mu}_{i,t-4}),\end{split}\end{equation}
for $i = 1, ..., N$.

		\section{Disaggregating Temporally Location Parameters}\label{ap:loc_disag}

 First, it is useful to start from the accounting relationship between monthly and quarterly flow variables; specifically, a quarterly variable in levels $Y_{i,t}$ must be equal to the three-month sum of its monthly sub-components $\tilde{Y}_{i,t-j}$, $j = 0,1,2$:
\begin{equation}
	\label{eq:accounting_constraint}
	Y_{i,t} =  \tilde{Y}_{i,t} + \tilde{Y}_{i,t-1} + \tilde{Y}_{i,t-2}.
\end{equation}
To account for heteroskedasticity and multiplicative components, however, the data are generally taken in logarithms, such that the variables modelled are $Y'_{i,t} = \text{log}Y_{i,t}$ and $\tilde{Y}'_{i,t} = \text{log}\tilde{Y}_{i,t}$. Since the sum of the logarithms is not equal to the logarithm of the sum, the accounting constraint takes a non-linear form:
\begin{equation}
	\label{eq:accounting_constraint_logs}
	Y'_{i,t} =  \text{log}\Bigl[\text{exp}\tilde{Y}'_{i,t} + \text{exp}\tilde{Y}'_{i,t-1} + \text{exp}\tilde{Y}'_{i,t-2}\Bigr].
\end{equation}

This type of non-linearity becomes difficult to handle once the first differences in logs are modelled. It is possible, however, to work from a linear approximation.~\citet{mitchell_2005} show that
\begin{equation}
	\label{martin_approx}
	\sum^2_{j=0}\text{h}(x_{t-j}) \approx 3 \ \text{h}\Biggl(\frac{\sum^2_{j=0}x_{t-j}}{3}\Biggr),
\end{equation}
where $h(.)$ is a non-linear transformation and $x_{t}$ a smooth variable. The approximation \eqref{martin_approx} is a second-order approximation because the first-order errors sum to zero. When monthly values in a quarter are close to the monthly average over the quarter, which is usually the case with seasonally adjusted figures, the approximation error introduced is negligible. Using this approximation equation, \eqref{eq:accounting_constraint_logs} can be written as
\begin{equation}
	\label{eq:accounting_constraint_ap}
	Y'_{i,t} = \text{log}3 + \frac{1}{3}\tilde{Y}'_{i,t} + \frac{1}{3}\tilde{Y}'_{i,t-1} + \frac{1}{3}\tilde{Y}'_{i,t-2}.
\end{equation}

Using approximation \eqref{eq:accounting_constraint_ap}, it is possible to write the quarter-on-quarter log difference $y_{i,t} = Y'_{i,t}-Y'_{i, t-3}$ as a function of the monthly log differences $\tilde{y}_{i, t} = \tilde{Y}'_{i, t}-\tilde{Y}'_{i, t-1}$, as
\begin{equation}
	\begin{split}
		y_{t,i} &= \frac{1}{3}\tilde{y}_{i,t}  + \frac{2}{3}\tilde{y}_{i,t-1} + \tilde{y}_{i,t-2} + \frac{2}{3}\tilde{y}_{i,t-3} + \frac{1}{3}\tilde{y}_{i,t-4}.
	\end{split}
\end{equation}
	
		\section{Simulation experiments}\label{sec:simulations}
		
		\subsection{The factor model as a data generating process}\label{ap:sim_dgp}
		
		This set of simulation investigates the path generated by the model when it is iterated forward which shades light on the behaviour of the model. It is also particularly relevant for multi-step ahead nowcasting, i.e. in the first two months of each quarter when missing observations need to be sampled. If the paths generated by the model are unrealistic (explosive) then this would affect the quality of the nowcasts and their associated uncertainties. 
		
\begin{figure}[H]
			\centering
			\includegraphics[width=0.9\textwidth]{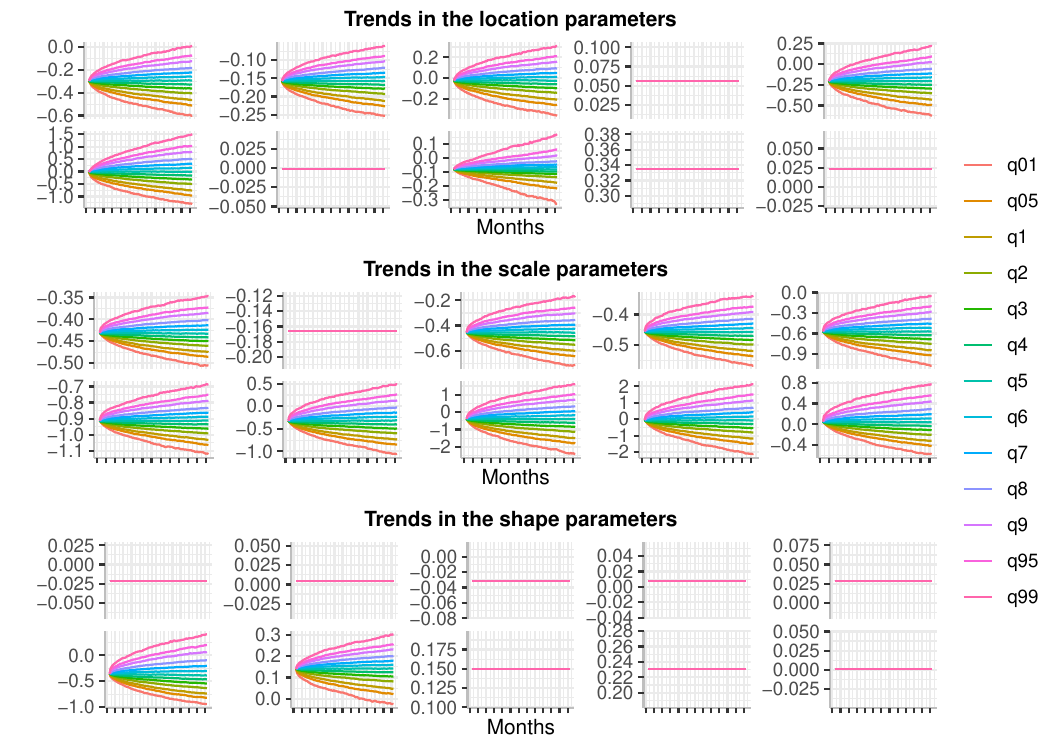}
			\vspace{-0pt}
			\caption{Quantiles of the generated paths from the trend components of the monthly factor model. Quantiles for the probabilities $0.01, 0.05, 0.1, 0.2, ..., 0.9, 0.95, 0.99$ are shown.The mixed-frequency factor model is first estimated with the latest available vintage and then iterated forward for ten years 1000 times.\label{fig:dgp_10y}}
		\end{figure}

		The experiment is designed as follows. First the monthly factor model \eqref{ref:observation_equation} - \eqref{ref:factor_model}  is estimated using the latest vintage; second, the recursion is re-started at the last period and iterated forward for ten years 1000 times. Figure \ref{fig:dgp_10y} shows sample quantiles from the resulting paths of the location, scale and shape trend components for each series of the model. The generated components are well-behaved without any explosive paths. Straight lines indicate that the corresponding component and estimated to be constant. The fact that the multi-step ahead nowcasts are well-calibrated (see section \ref{sec:results}) suggests also that the model is well-behaved. 
		 
		\subsection{Convergence of the filtered parameters}\label{ap:sim_filter}
\begin{figure}[H]
			\centering
			\includegraphics[width=1\textwidth]{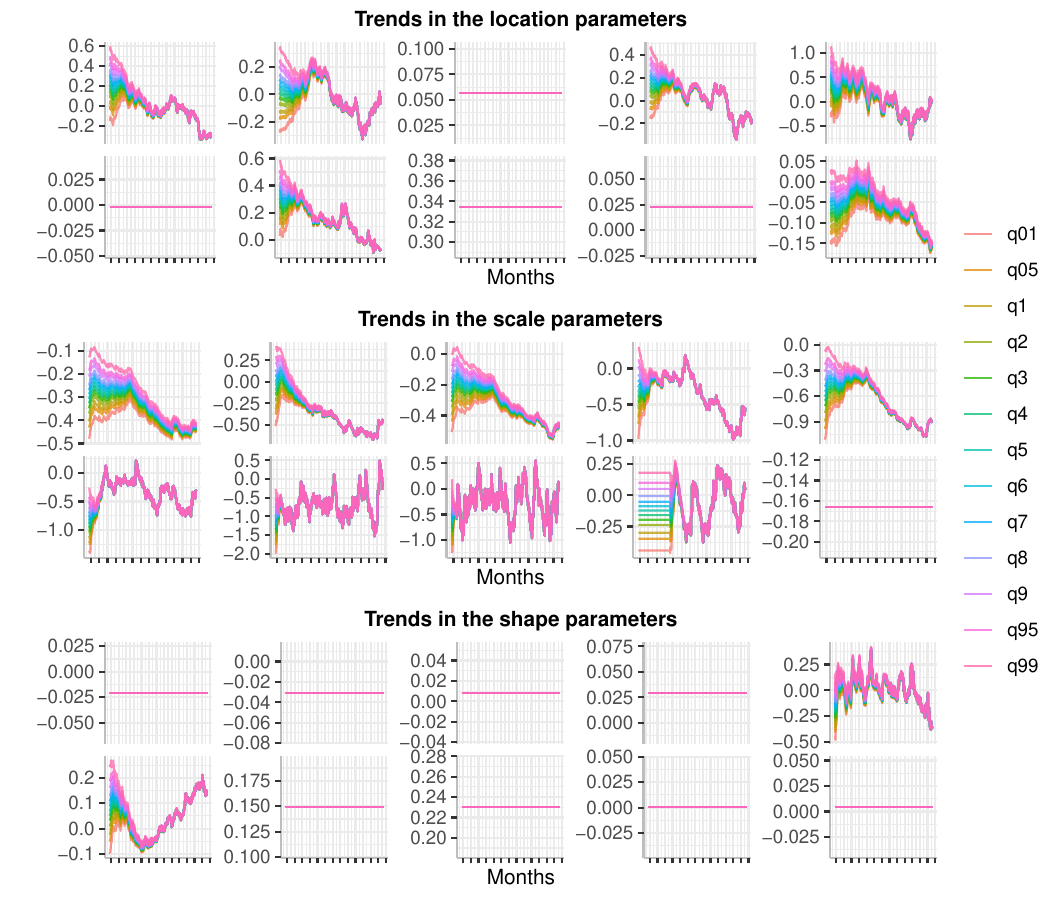}
			\vspace{-0pt}
			\caption{Quantiles from the paths of the filtered trend components for each of the series used in the monthly model. Quantiles for the probabilities $0.01, 0.05, 0.1, 0.2, ..., 0.9, 0.95, 0.99$ are shown. The parameters of the factor model are first estimated with the lastest vintage before adding noise to the initial values of the trends and iterating the filter forward. Results from 1000 simulations. \label{fig:filter_sim}}
		\end{figure}

		This experiment analyses the convergence of the filtered components following disturbances in their initial conditions. Simulations are generated as follows. The first step is similar to the previous experiment where the monthly factor model \eqref{ref:observation_equation} - \eqref{ref:factor_model}  is estimated using the latest vintage. Second, a random error is added to the initial values of the location, scale and shape trend components. The error follows a Normal distribution with mean zero and variance equal to the squared standard deviation of the trend component it affects. Third the filter is iterated forward. Figure \ref{fig:filter_sim} shows the results from 1000 simulations. It takes about a third of the total number of observations for the components to converge. A small amount of noise sometimes remains but overall the filter tends to be stable. Straight lines indicate that the corresponding component and estimated to be constant.

		\subsection{Assessing estimation stability through model-based noise}\label{ap:sim_estimation}
	
	The maximum likelihood approach for estimating score driven models has been studied thoroughly in \citet{blasques_2020}. However, there is no guarantee that the model with location, scale and shape common factors can be estimated precisely enough with the amount of data at hand. To investigate this issue, a set of simulations designed as follow is carried out. First, the monthly model with location, scale and shape common factors is estimated on the latest available vintage. Then, (i) the parameters and unobserved components estimated in the first step are used to draw prediction errors and thus generate pseudo observations; (ii) these generated observations are used to estimate the model and the resulting estimated common factors are stored. Steps (i) and (ii) are repeated one thousand times to analyse the variation around the estimated common factors. 
	
\begin{figure}[H]
		\centering
		\includegraphics[width=0.99\textwidth]{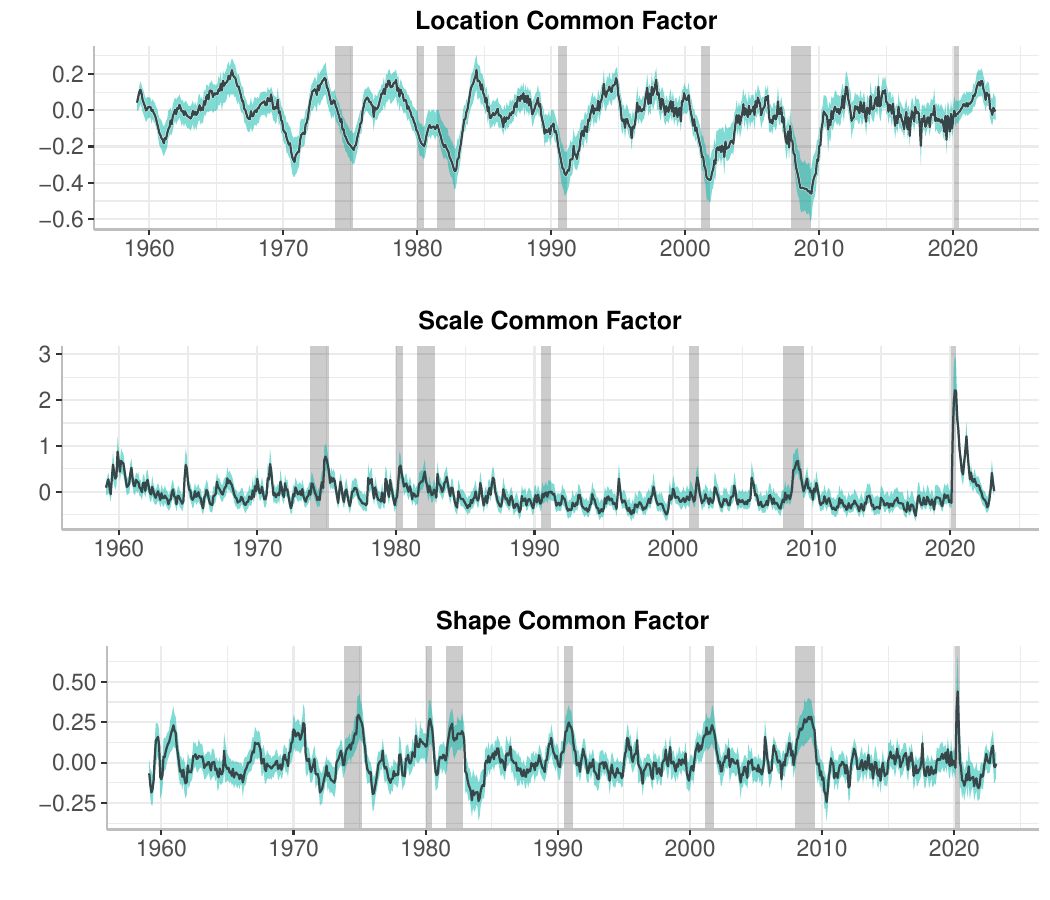}
		\vspace{-0pt}
		\caption{Mean and 68\% best critical region of the estimated common factors from 1000 simulations. The grey shaded areas indicate the recessions as classified by the NBER.\label{fig:sim}}
	\end{figure}

	Figure \ref{fig:sim} shows the mean estimates of the common factors with the 68\% best critical region. The estimation uncertainty is broadly similar for the shape and location and relatively higher for the scale. The average standard deviation from the factor estimates across the simulations is 0.11 for the shape, 0.09 for the location and 0.23 for the scale. This graph also shows that in-sample negative skewness increases during recessions and tend to increase early in the recovery before stabilising around zero. 
	
	\section{Accounting for missing data uncertainty when generating multi-step-ahead nowcasts}\label{ap:uncertainty}
	\subsubsection*{In the mixed-frequency factor model}
	If, for example, the target is quarterly GDP growth ending in month $t$, but real-time series are missing from $t-1$, then (i) the skew-t densities of these series at $t-1$ are used to draw prediction errors centred around their location estimates in $t-1$; (ii) the score driven recursion is applied separately on each prediction to retrieve sets of scale, shape and location parameters for period $t$; (iii) these new parameters yield skew-t densities which are used to draw prediction errors around the locations in $t$; (iv) each vector of prediction errors in $t$ yields a set of filtered parameters for GDP growth in $t$ through the filtering step of the score driven recursion; (v) these sets of parameters are used to draw predictions of GDP growth using the skew-t distribution; (vi) the density nowcast is eventually given by the empirical density attached to these GDP nowcasts.
	
	\subsubsection*{In the factor-augmented MIDAS model}
	Similarly, when using the two-step factor-augmented model it is important to account for the uncertainty attached to the common factors. The procedure is similar to the one outlined above but diverges in step (iv) where, instead of retrieving sets of parameters for GDP growth, the filtered estimates of the common factors are stored. The factor-augmented MIDAS model is estimated only once on the point estimates of these factors. Then the estimated model is run on all sets of common factors, where each sets yields predicted location, scale and shape parameters which are used to draw values of GDP growth with the skew t-distribution. Finally, like with the mixed-frequency factor model, the density nowcast is given by the empirical density attached to these GDP nowcasts.

	\end{appendices}

\end{document}